\documentclass[10pt,journal,cspaper,compsoc]{IEEEtran}
\usepackage{color}
\usepackage{algorithm}
\usepackage{algpseudocode}
\usepackage{cite}

\usepackage{array}
\usepackage{eqparbox}
\usepackage{stfloats}
\usepackage{url}
\usepackage{color}
\usepackage{multirow}
\usepackage{booktabs}
\usepackage{amsmath,amsfonts}
\usepackage{amssymb}
\usepackage[colorlinks]{hyperref}
\usepackage{subfigure}
\usepackage{caption2}
\usepackage{graphicx}
\usepackage{tikz}

\newcommand{\pgftextcircled}[1]{
    \setbox0=\hbox{#1}%
    \dimen0\wd0%
    \divide\dimen0 by 2%
    \begin{tikzpicture}[baseline=(a.base)]%
        \useasboundingbox (-\the\dimen0,0pt) rectangle (\the\dimen0,1pt);
        \node[circle,draw,outer sep=0pt,inner sep=0.1ex] (a) {#1};
    \end{tikzpicture}
}

\newcommand{\pgftextcircledblk}[1]{
    \setbox0=\hbox{#1}%
    \dimen0\wd0%
    \divide\dimen0 by 2%
    \begin{tikzpicture}[baseline=(a.base)]%
        \useasboundingbox (-\the\dimen0,0pt) rectangle (\the\dimen0,1pt);
        \node[circle,draw,outer sep=0pt,inner sep=0.1ex,fill=blue] (a) {#1};
    \end{tikzpicture}
}

\def\t0{{t_0}}

\def\N{{\mathbb N}}
\def\R{{\mathbb R}}        

\def\t0{{t_0}}

\def\Nm{{\mathcal N}}

\def\prox{{\rm prox}}

  \newtheorem{remark}{Remark}

\ifCLASSOPTIONcompsoc
\else
\fi
\ifCLASSINFOpdf
\else
\fi

\begin{document}
%
\title{From Social to Individuals: a Parsimonious Path of Multi-level Models for Crowdsourced Preference Aggregation}

\author{Qianqian~Xu,
        Jiechao~Xiong,
        Xiaochun~Cao*,
        Qingming~Huang*,~\IEEEmembership{IEEE Fellow}
        and~Yuan~Yao*  

\IEEEcompsocitemizethanks{\IEEEcompsocthanksitem Q. Xu is with the Key Laboratory of
Intelligent Information Processing, Institute of Computing Technology, Chinese
Academy of Sciences, Beijing 1000190, China, and also with the State Key Laboratory of Information Security (SKLOIS), Institute of Information Engineering, Chinese Academy of Sciences,
Beijing 100093, China, (email: qianqian.xu@vipl.ict.ac.cn).\protect\\
\IEEEcompsocthanksitem J. Xiong is with Tencent AI Lab, Shenzhen 518057, and also with the BICMR-LMAM-LMEQF-LMP, School of Mathematical Sciences, Peking University, Beijing 100871, China, (email: jcxiong@tencent.com). \protect\\
\IEEEcompsocthanksitem X. Cao is with State Key Laboratory of Information Security (SKLOIS), Institute of Information Engineering, Chinese Academy of Sciences,
Beijing 100093, China, (email: caoxiaochun@iie.ac.cn).\protect\\
\IEEEcompsocthanksitem Q. Huang is with the University of Chinese Academy of Sciences, and also with the Institute of Computing Technology of Chinese Academy of Sciences, Beijing 100190, China, (email: qmhuang@ucas.ac.cn).\protect\\
\IEEEcompsocthanksitem Y. Yao is with Department of Mathematics, and by courtesy, Computer Science and Engineering, Hong Kong University of Science and Technology, Hong Kong, (email: yuany@ust.hk). \protect\\

\IEEEcompsocthanksitem *Corresponding author.
}
}
%
%

\markboth{IEEE Transactions on Pattern Analysis and Machine Intelligence}%
{Shell \MakeLowercase{\textit{et al.}}: Bare Demo of IEEEtran.cls for Computer Society Journals}
%


\IEEEcompsoctitleabstractindextext{%
\begin{abstract}

In crowdsourced preference aggregation, it is often assumed that all the annotators are subject to a common preference or social utility function which generates their comparison behaviors in experiments. However, in reality annotators are subject to variations due to multi-criteria, abnormal, or a mixture of such behaviors. In this paper, we propose a parsimonious mixed-effects model, which takes into account both the fixed effect that the majority of annotators follows a common linear utility model, and the random effect that some annotators might deviate from the common significantly and exhibit strongly personalized preferences. The key algorithm in this paper establishes a dynamic path from the social utility to individual variations, with different levels of sparsity on personalization. The algorithm is based on the Linearized Bregman Iterations, which leads to easy parallel implementations to meet the need of large-scale data analysis. In this unified framework, three kinds of random utility models are presented, including the basic linear model with $L_2$ loss, Bradley-Terry model, and Thurstone-Mosteller model. The validity of these multi-level models are supported by experiments with both simulated and real-world datasets, which shows that the parsimonious multi-level models exhibit improvements in both interpretability and predictive precision compared with traditional HodgeRank.

\end{abstract}

}

\maketitle

\IEEEdisplaynotcompsoctitleabstractindextext

%
\IEEEpeerreviewmaketitle

\section{Introduction}

With the Internet and its associated explosive growth of information, individuals today in the world are facing with the
rapid expansion of multiple choices (e.g., which book to buy, which hotel to book, etc.). All of these examples yield comparisons without explicitly revealing an underlying preference or utility function. That is, only partial ordered choices subject to the preference are observed instead of the whole utility function, especially the paired comparisons that all partial ordered choices can be converted into. Therefore the aggregation of incomplete comparison data to reveal the global preference function has been one important topic in the last decades.

%

{
	\begin{figure}[b]
		\vspace*{-1cm}
		\footnotesize
		\begin{minipage}{\textwidth}\ \\
			\_\_\_\_\_\_\_\_\_\_\_\_\_\_\_\_\_\_\_\_\_\_\_\_\_\_\_\_\_\_\_\\
			A publicly available package (i.e. datasets and code) to reproduce\\ our experiments can be downloaded from\\
			https://github.com/qianqianxu010/PAMI2017

		\end{minipage}
	\end{figure}
}

In recent years, researchers in this odyssey usually take three approaches: (i) common consensus, that assumes all users' choices are stochastic revelation of a common global preference or utility function on candidates; (ii) collaborative filtering for personalized ranking, which often assumes different users have correlated preference functions represented by some low rank rating matrices; (iii) mixture of random utility models \cite{oh2014learning,farias2009data,negahban2012iterative}, that assumes the personal choice comes from one of a small set of underlying random utility models which are yet unknown.
Regarding common consensus pursuit, there has
been a large volume of studies from the social choice theory to modern ``rank aggregation" in computer science\cite{PageRank,de1781memoire,pairedbook,Yu12,Osher11_retinex,osting2013statistical,negahban2012iterative,dwork2001rank,Hodge}, on how to consistently aggregate the pairwise comparisons into a global consensus ranking that summarizes the preference of all users. On the other hand, low rank models of collaborative filtering \cite{salakhutdinov2008bayesian,rennie2005fast,yi2013inferring,lu2015individualized}
assume that there exist a small number of underlying intrinsic utility functions such that every individual's personalized preference is a linear combination
of these intrinsic utility functions, using nuclear norm as a penalty, while mixture models can be consistently recovered using tensor moment matching methods \cite{oh2014learning}. However, few work takes the wide spectrum by considering both the social preference and individual variations simultaneously.
In other words, they do not take
into account the multi-level hierarchies from social choice to
individuals.

\begin{figure}
\renewcommand{\captionfont}{\footnotesize \bfseries}
 \begin{center}
  \subfigure[Common preference with six representative group preferences]{
\includegraphics[width=0.45\linewidth]{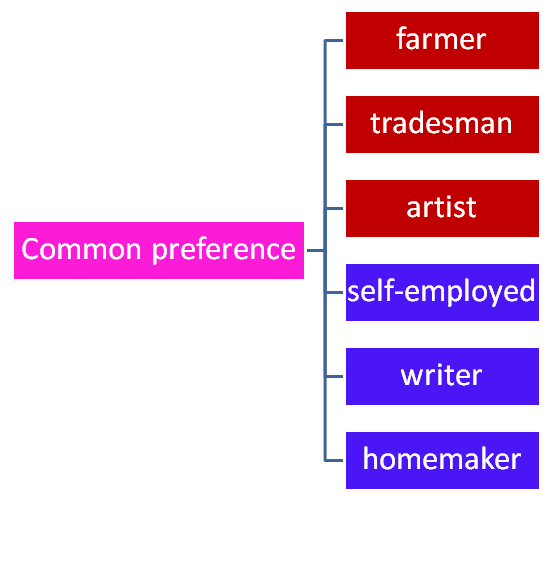}}
   \subfigure[Regularization paths of all 21 occupation group preferences]{
\includegraphics[width=0.5\linewidth]{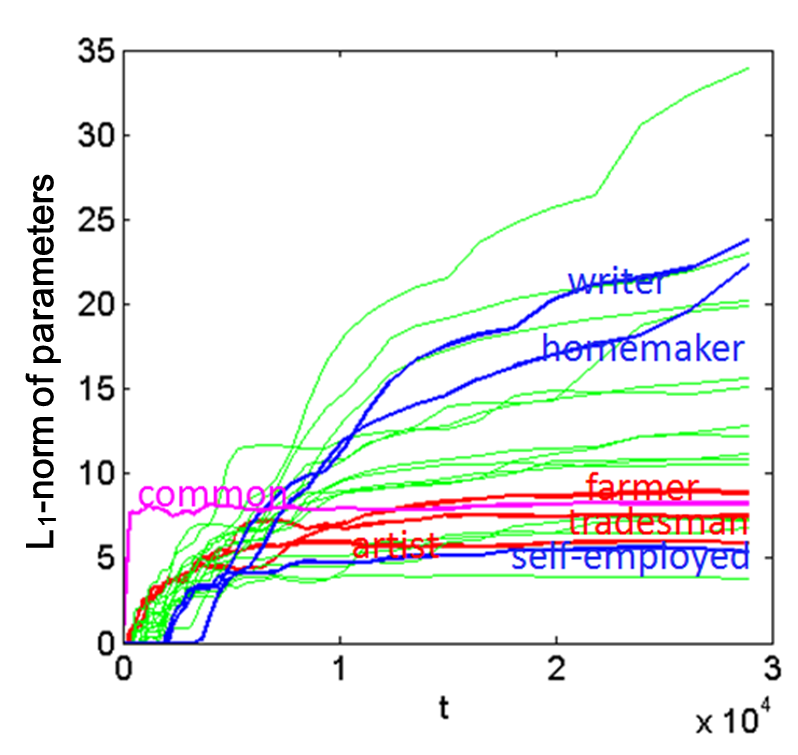}}
\caption{A two-level preference learning in MovieLens: (a) the common preference with six representative occupation group preference; (b) the purple is the common preference, the remaining 21 paths represent the occupation group preferences, the red are the three groups with most distinct preferences from the common, the blue are the three groups with most similar preferences to the common, and the green ones are the others.} \label{fig:moviepath}
\end{center}
\end{figure}

To see the nature of such hierarchies in preference learning, let's consider the movie rating from MovieLens dataset for example. Fig.\ref{fig:moviepath} shows a two-level movie preference functions learned from this dataset: the common preference and 21 group preferences. Fig.\ref{fig:moviepath} (a) illustrates this two-level hierarchical model with six representative groups, among which farmer, tradesman, artist are the top 3 groups exhibiting a large deviation from the common preferences, while self-employed, writer, homemaker are those showing similar preference with the common. Such results suggest that the main consuming groups of this website include homemaker, writer, and un-employed, who have more freedom to spend their time compared with other occupations. A coarse-grained model may just consider the common preference for all the users, or a refined model may incorporate these group variations to reflect diversity, while a further refined case may consider diversity in individual level. Fig.\ref{fig:moviepath} (b) shows the group preference diversity using the methodology proposed in this paper. The purple curve represents the common preference, while the remaining 21 curves there represent the 21 occupation group preferences in regularization paths, of which the earlier popping up to be nonzero, the more salient distinction is the group preference from the common. At different locations of $t$-axis, models of different diversity levels can be chosen.

In addition to the intrinsic preference diversity among users, there are abnormal behaviours of participants in crowdsourcing experiments due to diverse environment. Even they might share the same preference or utility function in making choices, they might suffer various disturbances during the experiments. For example, i) one typically clicks one side more often than another. As some pairs are highly confusing or annotators get too tired, in these cases, some annotators tend to click one side hoping to simply raise their record to receive more payment; while for pairs with substantial differences, they click as usual. ii) some extremely careless annotators, or robots pretending to be human annotators, actually do not look at the instances and click one side all the time to quickly receive payment for work. Such a kind of behavior is called the annotator's position bias which has been studied in \cite{day1969position}.

These examples above suggest us that we have to take into account of user or annotator specific variations in a crowdsourced preference aggregation task. In this paper, we propose \emph{a simple dynamic scheme that can learn multi-level utility models from the social common preference to individual diversity in a unified spectrum}, adapted to different statistical models (e.g. linear, Bradley-Terry, and Thurstone-Mosteller etc).

As the classical social choice theory \cite{Arrow51} points out, preference aggregation toward a global consensus is doomed to meet the conflicts of interests. \emph{What is a suitable way to quantitatively analyze the conflicts of interests}?

In this paper, we are inspired by the Hodge-theoretic approach proposed in \cite{Hodge} which decomposes the pairwise comparison data into three orthogonal components: the global consensus ranking, the local inconsistency as triangular cycles, and the global inconsistency as harmonic cycles.  Instead of merely extracting from the data the global ranking component, often called \emph{HodgeRank}, the latter two are both cycles, collectively decoding all the conflicts of interests in the data. To decipher the sources of the conflicts of interests, we further decompose the cycles by considering two types of annotator-specific variations here: annotator's personalized preference deviations from the common ranking which characterize multi-criteria in comparisons, and annotator's position bias which deteriorates the quality of data. This results in a linear mixed-effects extension of HodgeRank, called Mixed-Effects HodgeRank here. Such a principle can be applied to various generalized linear models that will be studied in this paper.



To initiate a task of crowdsourced preference aggregation, we usually assume the majority of participants share a common preference interest and behave rationally, while deviations from that exist but are sparse. So a parsimonious model is desired in this paper, with sparsity structure on personalized preference deviations and position biases. Due to the unknown amount of such sparse random effects in reality, it is natural to pursue a family of parsimonious models at a variety levels of sparsity. Algorithmically we developed the Linearized Bregman Iterations (LBI) as discretized Inverse Scale Space method in our setting, which is a simple iterative procedure generating a sequence of parsimonious models, evolving from the common global ranking in HodgeRank, to annotator's personalized ranking with a fully parametric model that might overfit the data. As the algorithm iterates, typically it appears early the large deviations in a personalized preference or abnormal behaviour, and the annotators who follow the common show at a later stage. In practice when the number of participants is large and sample size is relatively small, early stopping regularization is needed to prevent the overfitting in full model. Due to the algorithmic simplicity, it allows an easy (synchronized) parallel implementation to meet the need of large-scale data analysis.
%

As a summary, our main contributions in this new framework are highlighted as follows:

\begin{itemize}
\item[(A)] A linear mixed-effects extension of HodgeRank including both the fixed effect of common ranking, and the random effects such as annotator's preference deviations and position bias, which can be easily extended to generalized linear models including Bradley-Terry (BT) and Thurstone-Mosteller (TM) models that improve the efficiency for binary comparison data than the basic linear model (associated with the $L_2$ loss).
\item[(B)] A path of parsimonious estimates of the preference deviation and position bias at different sparsity levels, based on Linearized Bregman Iterations as a discretization of Inverse Scale Space method, which allows a simple synchronized parallelization for an almost linear speed-up.
\end{itemize}

This paper is an extension of our conference paper \cite{mm16}, where we proposed a basic linear mixed-effect model which not only can derive the common preference on population-level, but also can estimate an annotator's large preference/utility deviation in an individual-level, as well as an abnormal annotator's position bias. However, there are some limitations in this work.
First, it does not aim to predict the preferences (social or individual)
based on features of new users and/or new alternatives. Such
featured data are ubiquitous in E-commerce etc., such as
recommendations of books, movies, and restaurants, based
their styles and categories of users. To learn such preference
or ranking functions with predictive power on unseen products, a feature representation of the candidates in comparison must be used as model input in addition
to the local ranking orders. Second, other types of models are not studied, such as generalized linear models which are particularly efficient for discrete choice data.
In current new version, we propose \emph{a unified framework that includes various generalized linear models which learn both the social preference functions
based on features of alternatives to-be-compared and personalized utility functions conditioning on user categories}.
Such a model with at least two levels of diversity, enables us to simultaneously learn a
coarse-grained social social function together with fine-grained personalized rankings, equipped with prediction
power for the choices of new users on new alternatives. In this paper, we shall see that the Linearized Bregman Algorithm can be adapted to all these generalized linear models with fast and parallel path algorithms, and particularly enjoys the improved statistical precision of generalized linear models for binary comparisons in real world datasets, without losing the algorithmic simplicity in basic linear model.

The remainder of this paper is organized as follows. Sec.\ref{sec:relatedwork} contains a review of related works.
 Then we systematically introduce the methodology for parsimonious mixed-effects HodgeRank estimation in Sec.\ref{sec:methodology}. Extensive experimental validation based on one simulated and three real-world crowdsourced datasets are demonstrated in Sec.\ref{sec:experiment}. Finally, Sec.\ref{sec:conclusion} presents the conclusive remarks.

\section{Related Work} \label{sec:relatedwork}


Statistical preference aggregation, in particular ranking or rating from pairwise comparisons,
is a classical problem which can be traced back to the $18^{th}$ century.
Various methods have been studied for this problem, including the Borda
count \cite{de1781memoire}, maximum
likelihood method such as the Bradley-Terry model \cite{pairedbook}, rank centrality (PageRank/MC3) \cite{negahban2012,dwork2001rank}, and most recently, HodgeRank \cite{Hodge}.

HodgeRank, as an application of combinatorial Hodge theory to the preference or rank aggregation problem from pairwise comparison data, was first introduced in~\cite{Hodge}, inspiring a series of studies in computer science~\cite{mm11,hodge_l1,osting2013enhanced} and game theory~\cite{Parrilo11_gameflow}, in addition to traditional applications in fluid mechanics~\cite{Chorin93} and computer vision~\cite{Yuan09_hodge}, {etc}. It is a general framework to decompose paired comparison data on graphs, possibly imbalanced (where different candidate pairs may receive different number of comparisons) and incomplete (where every voter may only give partial comparisons), into three orthogonal components (gradients, local cycles, and harmonic cycles). In these components HodgeRank not only provides us a mean to determine a global ranking from paired comparison data under various statistical models (e.g., Uniform, Thurstone-Mosteller, Bradley-Terry, and Angular Transform), but also measures the inconsistency of the global ranking obtained. The inconsistency shows the validity of the ranking obtained and can be further studied in terms of its geometric scale, namely whether the inconsistency in the ranking data arises locally or globally. Local inconsistency can be fully characterized by triangular cycles, while global inconsistency involves cycles consisting nodes more than three (harmonic cycles), which may arise due to data incompleteness and once presented with a large component indicates some serious conflicts in ranking data.

In \cite{Tran16,ICML14}, it shows that under a natural statistical model, where pairwise comparisons
are drawn randomly and independently from some
underlying probability distribution, the rank centrality (PageRank) and
HodgeRank algorithms both converge to an optimal ranking under
a ``time-reversibility" condition. However, PageRank is only able to aggregate the pairwise comparisons
into a global ranking over the items. HodgeRank not only provides us a mean to determine a global ranking under
various statistical models, but also measures the inconsistency of the global ranking obtained. Exploiting the random graphs,
we can efficiently control the global inconsistency via topology of random clique complexes \cite{tmm12} as well as the sampling efficiency \cite{OXXY16}.

However, all of these methods have a major drawback: they aim to find \emph{one} global
ranking thus cannot analyze the conflicts of interests or discrepancies across users. In HodgeRank \cite{Hodge}, such conflicts are encoded in the components of cyclic rankings, which are not user-specific. On the other hand, in crowdsourcing scenarios, users may vote following multi-criteria or under different environments that contribute to the preferential diversity. Deciphering such behaviors becomes necessary for a better exploit of crowdsourcing data.

Recently, some personalized ranking methods arose from the standard collaborative filtering (CF) approach that is based on matrix factorization \cite{salakhutdinov2008bayesian,rennie2005fast,yi2013inferring}. The key idea behind them is to find a low rank user rating matrix via nuclear norm regularization such that every user's utility is a linear combination of such low-rank ratings. However such models are not a natural fit in crowdsourcing scenarios where the majority of voters share some common preference while some annotators might deviate from that significantly. 

Beyond the CF approach, there are various techniques to model annotators' abnormal behaviors in general crowdsourcing \cite{Reviewer2_1,Reviewer2_2,Reviewer2_3,Reviewer2_4,Reviewer2_5,Reviewer2_6,Reviewer2_7,Reviewer2_8,Reviewer3_3,Reviewer3_2,Reviewer3_1}, etc. The basic idea
of these work is to characterize user quality using some probabilistic behavior models. The models roughly lie in two categories \cite{Reviewer3_3}: either a single parameter is associated with each user's quality indicating the probability that the annotator
correctly answers a task \cite{49,74}, or a general confusion matrix is used for each user as extensions from the classic work of Dawid and Skene (DS) \cite{dawid1979,RaykarYZJFVBM09,WhitehillRWBM09}. In particular, \cite{Reviewer3_1} considers task dependent user quality parameters or confusion matrices such that the majority follows the common parameter while some may deviate from that with personalized parameters; on the other hand, \cite{Reviewer3_2} directly exploits the correlations between user confusion matrices to discover hidden groups of users.

While these methods can model the quality of the workers in general crowdsourcing experiments for label aggregation, they lack the consideration for peculiarity in crowdsourced preference aggregation where every user may vote following some utilities. For example, in pairwise comparisons, the confusion matrix approach will lead to an adversarial mixture ranking model \cite{7835206}, where every voter follows a mixture of rational behavior by voting according to the common ranking model and abnormal behavior by voting according to its adversarial ranking. However, voters are not necessarily adversarial; for example, robot clickers on one side can be captured by position bias in our model and random clickers can be captured by his/her deviations in personalized ranking, both of which are clearly not adversarial voters. Therefore the models with quality parameter or confusion matrix above are coarse-grained models in crowdsourced ranking, insufficient to capture the preferential diversity. In this paper, we are inspired by the HodgeRank approach, and propose a parsimonious multilevel model for personalized rankings that decipher conflicts of interests but are not necessary adversarial, so may capture a wider or more refined preferential diversity in crowdsourced rank aggregation than previous models.

\section{Methodology}
\label{sec:methodology}

In this section, we systematically introduce the methodology
for parsimonious mixed-effects HodgeRank estimation. Specifically,
we first start from introducing the proposed mixed-effects model based on HodgeRank, in which three kinds of random utility models are presented including
the basic linear model with $L_2$ loss, Bradley-Terry model, and Thurstone-Mosteller model, etc.
Then we present a simple iterative algorithm called Linearized Bregman Iterations to generate paths of parsimonious models at different sparsity levels, followed by Synchronized Parallel LBI to meet the need
of large-scale data analysis. Finally, early stopping regularization is discussed in the end of this section.

\subsection{Mixed-Effects HodgeRank on Graphs}

Suppose there are $n$ alternatives or items to be ranked, represented by $n$ data points
with a feature matrix $\Phi=[\phi_{i}^T]_{i=1}^n \in \mathbb{R}^{n \times d}$, where $\phi_{i}$ is a $d$-dimensional feature vector representing item $i$. The pairwise comparison
labels collected from users
can be naturally represented as a directed comparison
graph $G = (V;E)$. Let $V = \{1,2,\dots,n\}$ be the vertex set of $n$ items and $E = \{(u,i,j): i,j\in V, u \in U\}$ be the set of edges, where $U$ is the
set of all users who compared items. User $u$ provides his/her preference between choice $i$ and $j$,  such that $y_{ij}^u>0$ means $u$
prefers $i$ to $j$ and $y_{ij}^{u}\leq 0$ otherwise. Hence we may assume $y: E\rightarrow R$ with skew-symmetry
(orientation) $y_{ij}^u=-y_{ji}^u$. The magnitude of $y_{ij}^u$ can represent the degree of preference and it varies
in applications. The simplest setting is the binary choice, where $y_{ij}^u = 1$ if $u$ prefers $i$ to $j$ and $y_{ij}^u=-1$ otherwise. In applications, users are often categorized by their classifications, such as occupations and ages, hence $y_{ij}^u$ may be a summary statistics of all the pairwise comparisons between $i$ and $j$ among the same category of users.

The general purpose of preference aggregation is to look for a global score $\theta\colon V\to \R$ such that

\begin{equation} \label{eq:ho_rank0}
\min_{\theta\in {\mathbb{R}}^{|V|}} L(\theta):= \sum_{i,j,u} \omega_{ij}^u \l(\theta_i - \theta_j, y_{ij}^u),
\end{equation}
where $\l(a,b)\colon \R\times \R\to \R$ is a loss function, $\omega_{ij}^u$ denotes the confidence weights on $\{i,j\}$ made by rater $u$ (for simplicity, assumed to be $\omega_{ij}^u=1$ for the provided voting data), and $\theta_i$ ($\theta_j$) represents the global ranking score of item \emph{i} (\emph{j}, respectively).
In HodgeRank, one benefits from the use of square loss $\l(a,b)=(a-b)^2$ which leads to fast algorithms to find optimal global ranking $\theta$, which becomes one component of a general orthogonal decomposition of paired comparison data \cite{Hodge}, i.e.
\[ y= global\ ranking \oplus cycles, \]
where the component \emph{cycles} can be further decomposed into
\[cycles = local\ cycles \oplus global\ cycles. \]

Local cycles are triangular cycles, e.g. $i \succ j \succ k\succ i $; while global cycles, also called harmonic cycles, are loops involving nodes more than three (e.g. $i \succ j \succ k \succ...\succ i$) and typically traversing all nodes in the graph. These cycles may arise due to conflicts of interests in ranking data. Therefore to analyze the statistical models of cycles is crucial to understand the conflicts of interests.

In crowdsourcing scenarios, the conflicts of interests are mainly due to two kinds of sources: the multi-criteria adopted by different annotators when they compare items in $V$; the abnormal behavior of annotators in the experiments, e.g. simply clicking one side of the pair when they got bored, tired, or distracted. From this viewpoint, the source of such cycles in HodgeRank are usually caused by the personalized ranking, position bias, and stochastic noise.

To be specific, together with the global ranking component in HodgeRank, we consider the following linear mixed-effects model for annotator's pairwise ranking:
\begin{equation}
y_{ij}^u \sim F( (\phi_i^T\eta+\phi_i^T\xi^u) - (\phi_j^T\eta+\phi_j^T\xi^u) + \gamma^u),
\end{equation}

\begin{itemize}
\item $\eta$ is the common preference parameter such that the inner product with the $i^{th}$ feature $\phi_i$, $\theta_i:=\phi_i^T\eta$ gives the common preference score on item $i$, as a fixed effect;
\item $\xi^u$ is the user's preference deviation parameter from the common consensus such that $\theta_i^u:=\phi_i^T(\eta+\xi^u)$ becomes user $u$'s personalized preference score, as a random effect;
\item $\gamma^u$ is an annotator's position bias, which captures the careless behavior by clicking one side during the comparisons;
\item The distribution $F$ can be arbitrary cumulative distribution function.
\end{itemize}

Here $\eta$ is population-level parameter which indicates some common coefficient weight vector of the feature. In reality, as the preference vary greatly
across different types of users, we allow each type of user to have their personalized
parameters. These personalized parameters can be obtained by adding some
random effects $\xi^u$ to the population parameter $\eta$, representing personalized deviations from the population behavior.
Moreover, $\gamma^u$ measures an annotator's position bias, i.e. the tendency of $u$ always clicking one side in paired comparison experiments. Under the random design of pairwise comparison experiments, a candidate should be placed on the left or the right randomly, so the position should not affect the choice of a careful annotator. However, some annotator might get confused, tired or distracted in experiments, such that he/she always clicks one side during some periods in experiments, which can be detected by such $\gamma^u\neq 0$ \cite{xu2016false}.

Considering the variety of applications, this model may incorporate several types of feature matrices $\Phi=[\phi_i^T]_{i\in V}$, motivated but are not limited to the following examples.

\begin{itemize}
\item \textbf{$\Phi$ is an identity matrix.} For example, in world-college ranking, $V$ consists of colleges to be ranked, and $y_{ij}^u=1$ indicates user $u$ prefers college $i$ to $j$. In this scenario,
we do not have the features of each college but only the pairwise comparisons obtained from users.
\item \textbf{$\Phi$ is low-level (or deep) visual features.} For example, in music ratings, $\phi_{i}$ can be the low-level audio features extracted from each audio frame (spectrum power, Zero Crossing Rate, intensity, bandwidth, pitch and MFCC, etc).
\item \textbf{$\Phi$ is categorical type.} For example, in movie ratings, $\phi_{i}$ can be the genres of movie $i$, (e.g., Action, Adventure, Animation, Comedy, Drama, etc). Or in dining restaurant ratings, $\phi_{i}$ can be the cuisine types (e.g., Bar, Cafe-Coffee-Shop, Cafeteria, Fast-Food, etc) of the restaurant $i$.
\end{itemize}

To make the notation clear, let $d^u\in \R^{|E| \times |V|}$ satisfies $d^u\theta(v,i,j) = 1_{(u=v)}(\theta_i-\theta_j)$ and $d = \sum_u d^u$. Let $A \in \R^{|E| \times |U|}$ satisfies $A\gamma(u,i,j) = \gamma^u$. Denote $\theta = \Phi\eta$, $\delta^u = \Phi\xi^u$, and $\beta=[\xi,\gamma]$. Let $X = [d^1\Phi,\dots, d^{|U|}\Phi,A]$, so $X\beta = \sum_u d^u \Phi \xi^u + A\gamma$.

Different distribution functions $F$ respond to different statistic models. For example, when $F$ is normal function or sub-gaussian function, it indicates data follows the normal distribution or sub-gaussian distribution, which means
\begin{equation} \label{eq:linear}
y_{ij}^u = (\phi_i^T\eta+\phi_i^T\xi^u) - (\phi_j^T\eta+\phi_j^T\xi^u) + \gamma^u + \varepsilon_{ij}^u,
\end{equation}
or in matrix form
\begin{equation} \label{eq:linear-matrix}
y = d\Phi\eta +  X\beta + \varepsilon.
\end{equation}
where $\varepsilon_{ij}^u$ measures the random noise in sampling which is of zero mean and bounded. For notational simplicity, we abuse the notation $y$ to denote the vector $[y_{ij}^u]$. In this case, the loss function is often the $L_2$ loss, the negative log-likelihood of Gaussian:
\begin{equation}\label{eq-loss}
L(\eta,\beta) = \frac{1}{2m}\|y - (d\Phi\eta + X\beta)\|_2^2.
\end{equation}
For robust statistics, one can also adopt $L_1$ loss \cite{osting2013statistical} or Huber's loss which is equivalent to the $L_2$ loss with sample-wise sparse $\gamma^u$.


For binary comparison data $y^u_{ij}\in \{\pm 1\}$, there is a family of generalized linear model (GLM) in statistics:
\begin{align}\label{eq:glm}
P(y^u_{ij} = 1) &= 1-P(y^u_{ij}=-1) \nonumber\\
&= \Psi((\phi_i^T\eta+\phi_i^T\xi^u) - (\phi_j^T\eta+\phi_j^T\xi^u) + \gamma^u)
\end{align}
or
\begin{equation}
\Psi^{-1}(P(y=1)) = d\Phi\eta + X\beta.
\end{equation}
where $\Psi(t)$ is a symmetric cumulative distribution function (CDF) whose continuous inverse is well-defined. For example, 

 1. \emph{Bradley-Terry} model:
\begin{align}
\Psi(t) = \frac{1}{1+e^{-t}}.
\end{align}

\begin{figure}
\renewcommand{\captionfont}{\footnotesize \bfseries}
 \begin{center}
   \subfigure[Linear]{
\includegraphics[width=0.2\textwidth]{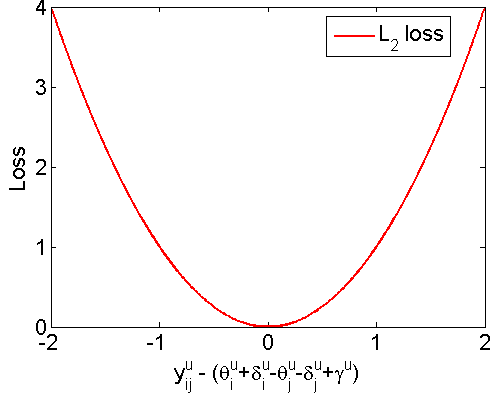}}
 \subfigure[BT \& TM]{
\includegraphics[width=0.19\textwidth]{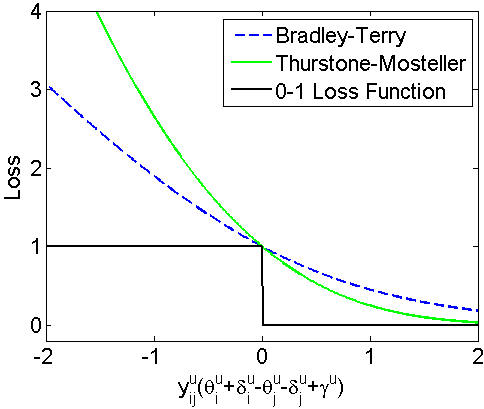}}
  \caption{Comparison of three loss functions.} \label{threeloss}
\end{center}
\end{figure}

 2. \emph{Thurstone-Mosteller} model:
 \begin{align}
 \Psi(t) = \frac{1}{\sqrt{2\pi}}\int_{-\infty}^{t}e^{-x^2/2}dx
 \end{align}
More models can be found in \cite{pairedbook,Hodge,mm11,tmm12}. We note that in general Hodge theoretical framework for binary pairwise comparison data, one can map binary comparison data into skew-symmetric flows on graphs by $\hat{y}=\Psi^{-1}(P(y^u_{ij}=1))$ and Hodge decomposition can be applied to such flows \cite{Hodge,tmm12}. In this paper, for the GLM model \eqref{eq:glm}, the loss function is chosen as the following negative log-likelihood,
\begin{equation}\label{eq:likelihood}
L(\eta,\beta) = -\frac{1}{m}\sum_{i,j,u}\log\Psi(y^u_{ij}(\phi_i^T(\eta_i+\xi_i^u) - \phi_j^T(\eta_j+\xi_j^u) + \gamma^u)).
\end{equation}
Here we use the symmetry $\Psi(-t) = 1-\Psi(t)$. Fig.\ref{threeloss} illustrates the comparisons of three losses, including $L_2$, Bradley-Terry (BT), and Thurstone-Mosteller (TM), respectively. One can see that as in classifications, Bradley-Terry and Thurstone-Mosteller provide convex surrogates \cite{BJM06} of binary comparison 0-1 loss with a better approximation than the L2 loss. Therefore one should expect that these two models may provide a better efficiency in reducing the pairwise mismatch (Kendall $\tau$-distance) from the observed data, as we shall see later in this paper.

\subsection{Parsimonious Paths of Multi-level Models with Linearized Bregman Iteration}
In crowdsourced preference aggregation scenarios with good controls, it is natural to assume a parsimonious model. In such a model, the majority of annotators carefully follows the common behavior governed by the fixed effect parameter $\theta$, while only a small set of annotators might have nonzero personalized deviations and abnormal behavior in position bias. This amounts to assume that parameter $\delta^u$ to be group sparse, i.e. $\delta_i^u$ vanishes for all $i$ simultaneously, and $\gamma^u$ to be sparse as well, i.e. zero for most of careful annotators.

Let's consider two representative scenarios:

$\bullet$ When $\Phi$ is an identity matrix, $\delta^u = \xi^u$, So such a sparsity pattern motivates us to consider the following penalty function with a mixture of LASSO ($L_1$) penalty on $\gamma$ and group LASSO penalty on $\xi^u$:
\begin{equation}\label{eq-penalty}
P(\beta) = \|\gamma\|_1 + \sum_{u}\|\xi^u\|_2.
\end{equation}
{\begin{remark}
Usually a normalization factor $\sqrt{n}$ is used before a group lasso penalty $\|\xi^u\|_2$, where $n$ is the group size of $\xi^u$. But here all the $\xi^u$ have the same group size, and $\|d^u\|_F = \sqrt{2}\|A^u\|_F$, so the column norm of $d^u$ is on average $\frac{\sqrt{2}}{\sqrt{n}}$ times of $\|A^u\|_F$, this basically cancels out the factor $\sqrt{n}$. So here we just use this simple formula.
\end{remark}
\medskip

$\bullet$ When $\Phi$ is low-level (or deep) visual features, such a sparsity pattern only needs to assume traditional LASSO ($L_1$) penalty on both $\gamma$ and $\xi^u$.
\begin{equation}\label{eq-penalty}
P(\beta) = \|\gamma\|_1 + \sum_{u}\|\xi^u\|_1.
\end{equation}

Given the Loss function and Penalty function, the following Linearized Bregman Iterations (LBI) give rise to a sequence of parsimonious (sparse) models:
\begin{subequations}\label{eq:lbi0}
\begin{align}
\eta^{k+1} & = \eta^k - \alpha \kappa  \nabla_\eta L(\eta^k,\beta^k) \label{eq:lbi0-a}\\
z^{k+1} & = z^k - \alpha \nabla_\beta L(\eta^k,\beta^k), \label{eq:lbi0-b}\\
 \beta^{k+1} &=\kappa \cdot {\prox}_{P}(z^{k+1}), \label{eq:lbi0-c}
\end{align}
\end{subequations}
where $\beta^0 = 0$, $\eta^0 = 0$, $z^0=0$, $k$ is the iteration index, and the proximal map associated with the penalty function $P$ is given by
\[ \prox_P(z) = \arg\min_{v \in R^{(|V|+1)|U|}} \left( \frac{1}{2}\| v - z\|^2 + P(z) \right ). \]
Here variable $z$ is an auxiliary parameter used for gradient descent, where by Moreau decomposition $z = \rho+\beta/\kappa, \rho \in \partial P(\beta)$.

The Linearized Bregman Iteration \eqref{eq:lbi0} generates a path of global ranking score estimators $\theta^k = \Phi\eta^k$ and sparse estimators for preference deviation and position bias, $\beta^k=(\xi^k,\gamma^k)$. It starts from the null model, and evolves into parsimonious mixed effect models with different levels of sparsity until the full model, often overfitted. To avoid the overfitting, early stopping regularization is required to find an optimal tradeoff between the model complexity and in-sample error. In this paper, we find that cross validation works to find the early stopping time that will be discussed in Sec.\ref{sec:cv}. 

The Linearized Bregman algorithm was firstly introduced in \cite{OBG+05} as a scalable algorithm for large scale image restoration with TV-regularization. It has several advantages than the widely used LASSO-type convex regularizations. First of all, it is simpler than LASSO in generating the sparse regularization paths: instead of a parallel run of several optimization problem over a grid of regularization parameters, a single run of LBI generates the whole regularization path. LBI is thus desired in dealing with big problems.

The main advantage of such a three line algorithm, not only lies in its algorithmic simplicity, but also gives us more statistical precision.
In fact, it has been shown \cite{osher2014} that LBI can be less biased than LASSO as if nonconvex regularizations \cite{FanLi01}. Precisely as $\kappa\to \infty$ and $\alpha_t\to 0$, the limit dynamics of Linearized Bregman Iterations in sparse linear regression may achieve the model selection consistency under nearly the same condition as LASSO yet return the unbiased Oracle estimator, while the LASSO estimator is well-known biased. In our case, LBI \eqref{eq:lbi0} is a discretization of the following limit differential inclusion:
\begin{subequations}\label{eq:iss}
\begin{align}
\frac{d \eta}{d t} & =-  \nabla_\eta L(\eta,\beta) \label{eq:iss-a}\\
\frac{d \rho}{d t} & = - \nabla_\beta L(\eta,\beta), \label{eq:iss-b}\\
\rho(t) &\in \partial P(\beta(t)). \label{eq:iss-c}
\end{align}
\end{subequations}
It evolves as gradient descent flows on a subspace restricted by $\rho(t) \in \partial P(\beta(t))$. For example, for a LASSO penalty $P(\beta)=\|\beta\|_1$, the support set $S_t = \{i: \beta_i(t)\neq 0\}$ must lead to $\rho_{S_t}(t)=\pm 1$ as a constant function, which leads to $\nabla_\beta L(\eta,\beta_{S_t}(t))=-\frac{d \rho_{S_t}}{dt}=0$, whence $\beta(t)$ is a minimizer (maximum likelihood estimator) restricted on the support set $S_t$. Such a minimizer is unbiased when sign consistency is reached, hence is statistically more accurate than any convex regularized estimator such as LASSO. For more details, we refer the readers to see \cite{osher2014} and references therein. Dynamics \eqref{eq:iss} is often called \emph{Inverse Scale Space} as it evolves with coarse-to-fine models, where at different $t$ one obtains models at different levels.



Here we give some remarks on the implementation details of the Linearized Bregman Iterations \eqref{eq:lbi0}.
\begin{itemize}
\item The parameter $\kappa$ determines the bias of the sparse estimators, a bigger $\kappa$ leading to the less biased ones. The parameter $\alpha$ is the step size which determines the precise of the path, with a large $\alpha$ rapidly traversing a coarse-grained path. However one has to keep $\alpha \kappa$ small to avoid possible oscillations of the paths, e.g. $\alpha \kappa \|d\Phi\Phi^Td^T+XX^T\|_2/m<2$. The default choice in this paper is $\alpha  = \frac{m}{\kappa\|d\Phi\Phi^Td^T+XX^T\|_2}$ as a tradeoff between performance and computation cost.
\item The step \eqref{eq:lbi0-a} can also be replaced by $$\eta^{k+1}  = \arg\min_{\theta}L(\eta,\beta^k)$$ if it is easy to solve.
\item Now we turn to simplify the third step \eqref{eq:lbi0-c} with an explicit formula for the proximal map with the particular penalty function defined in Eq. (\ref{eq-penalty}). Recovering $\beta^{k+1}$ from $z^{k+1}$ is equivalent to the following group shrinkage on each group component of $\beta$, i.e. $\gamma^u$ and $\xi^u$:
\begin{eqnarray}
\beta^{k+1} &=& \kappa \mathbf{Shrinkage}(z^{k+1})\\
&\triangleq &\left\{
\begin{aligned}
\xi^{u,k+1} &= \kappa\max(0,1-1/\|z_{\xi^u}\|_2)z_{\xi^u} \nonumber \\
\gamma^{u,k+1} &= \kappa\max(0,1-1/|z_{\gamma^u}|)z_{\gamma^u} \nonumber \\
\end{aligned}
\right.
\end{eqnarray}
\end{itemize}


Now we are ready to give the following Linearized Bregman Algorithm for our Mixed-Effects HodgeRank as Alg. \ref{alg-LBI-LME}.
\begin{algorithm}
\small
\caption{LBI for ME-Model}\label{alg-LBI-LME}
\textbf{Input:} Data $(d,X,y)$, damping factor $\kappa$, step size $\alpha$.\\
\textbf{Initialize:} $\beta^0 = 0,\eta^0 = 0,z^0=0,t^0=0$.\\
{\textbf{for $k=0,\dots,K$ do}
\begin{enumerate}
\item $pred^k = d\Phi\eta^k + X\beta^k$
\item $g^{k+1} = \mathbf{Gradient}(y,pred^k)$
\item $\eta^{k+1} = \eta^k - \frac{\alpha\kappa}{m}\Phi^Td^Tg^{k+1}$ \label{alg2-step1}
\item $z^{k+1}  =z^{k} - \frac{\alpha}{m} X^Tg^{k+1}$\label{alg2-step2}
\item $\beta^{k+1} = \kappa\mathbf{Shrinkage}(z^{k+1})$
\item $t^{k+1} = (k+1)\alpha$.
\end{enumerate}
\textbf{end for}}\\
\textbf{Output:} Solution path $\{t^k, \eta^k,\beta^k\}_{k= 0,1,\dots,K}$.
\end{algorithm}

The $\mathbf{Gradient}$ function is different for different models. For linear model
$$\mathbf{Gradient}(y,pred) = pred - y.$$
While for GLM, it can be written as follows:
$$\mathbf{Gradient}(y,pred) = - \psi(y.*pred).*y./\Psi(y.*pred)$$
Here $.*$ and $./$ means entry-wise multiplication/division, respectively, and $\psi(t) = \Psi'(t)$ is the probability density function corresponding to $\Psi(t)$. Here $\psi(t) =  \frac{e^t}{(1+e^{t})^2}$ corresponds to Bradley-Terry model and $\psi(t) = \frac{1}{\sqrt{2\pi}}e^{-x^2/2}$ for Thurstone-Mosteller model.

\subsection{Synchronized Parallel LBI}
\label{sec:synpar}

To meet the needs of large-scale data analysis, we would like to introduce a vanilla version of synchronized parallel LBI.
The algorithm \ref{alg-LBI-LME} only needs matrix-vector multiplication, which is easy to be parallelized. Algorithm \ref{alg:LB-parallel} is the synchronized parallel version of Algorithm \ref{alg-LBI-LME}.

\begin{algorithm}
\raggedright
{\small{
\caption{SynPar-LBI of Algorithm \ref{alg-LBI-LME}}\label{alg:LB-parallel}
\textbf{Initialization:} Given parameter $\kappa$, $\triangle t$ and thread number $P$, $k=0, z^0=0, w^0=0$.\\
\textbf{Split data and variables:} $U = \bigcup_{i=1}^P U_i$,~\,$\{1,\dots,p\} = \bigcup_{i=1}^P J_i$.\\
\textbf{Iteration:} For each thread $i$\\
 ~\, $update_i = 0$. For all $u$ in $U_i$,
\begin{subequations}
\begin{align}
pred^k_{u} &= d^u\Phi\eta^k + d^u\Phi(\xi^u)^k + A^u\gamma^u.\\
g^{k+1}_{u} &= \mathbf{Gradient}(y^u,pred^k_{u})\\
z_{\xi^u}^{k+1} &= z_{\xi^u}^k +  \frac{\alpha}{m} \Phi^T(d^u)^T g^{k+1}.\\
z_{\gamma^u}^{k+1} &= z_{\gamma^u}^k +  \frac{\alpha}{m} (A^u)^T g^{k+1}.\\
(\xi^u)^{k+1}&=\kappa\,\mathrm{shrink}(z_{\xi^u}^{k+1}).\\
(\gamma^u)^{k+1}&=\kappa\,\mathrm{shrink}(z_{\gamma^u}^{k+1}).\\
update_i &= update_i + \Phi^T(d^u)^T g^{k+1}.
\end{align}
\end{subequations}
 ~\, Synchronize.\\
$$\eta^{k+1}_{J_i} = \eta^{k}_{J_i} + \frac{\alpha\kappa}{m} \sum_{j=1}^P update_{j}.$$
 ~\, Synchronize.\\
\textbf{Stopping:} exit when stopping rules are met. 
}}
\end{algorithm}

\subsection{Early Stopping Regularization} \label{sec:cv}
The Alg.\ref{alg-LBI-LME} or  \ref{alg:LB-parallel} actually returns a solution path with many estimators of different sparsity. So we need to find an optimal stopping time among $t^k=\alpha k$ to choose some best estimators and avoid overfitting. Here we sketch the procedure of cross-validation to choose the optimal stopping time:
\begin{itemize}
\item Given the training data, fix $\kappa$ and $\alpha$, then split the data into $K$ folds. Then choose a list of parameter $t$.
\item \textbf{for $k=1,\dots,K$ do}
	\begin{enumerate}
	\item Run Alg.\ref{alg-LBI-LME} or \ref{alg:LB-parallel} on the training data except $k$-th fold to get the solution path.
	\item For pre-decided parameter list of $t$, use a linear interpolation to get $(\eta(t),\beta(t))$.
	\item On the $k$-th fold of training data, use the estimator $(\eta(t),\beta(t))$ to predict, and then compute prediction error.
	\end{enumerate}
	\textbf{end for}
\item  Return the optimal $t_{cv}$ with minimal average prediction error.
\end{itemize}
\textbf{Remark:} Because the Alg.\ref{alg-LBI-LME} or \ref{alg:LB-parallel} only return the estimator at discrete $\{t^k\}$ and may not contain the pre-decided parameter $t$, so we use a linear interpolation of the nearest two estimator $(\eta^{k},z^{k})$ and $(\eta^{k+1},z^{k+1})$ to approximate $(\eta(t),z(t))$. $\beta(t)$ is further obtained by using $\mathbf{Shrinkage}(z(t))$.

\section{Experiments}\label{sec:experiment}

In this section, four examples are exhibited with both
simulated and real-world data to illustrate the validity of
the analysis above and applications of the methodology proposed.
The first example is with simulated data while the
latter three exploit real-world data collected by crowdsourcing.

\subsection{Simulated Study}

\textbf{Settings} We validate the proposed algorithm on simulated data with $n=|V|=20$ labeled by 100 users.
Specifically, we first generate the feature matrix for each nodes: $\Phi=[\phi_{i}^T]_{i=1}^n \in \R^{n \times d}$, where $\phi_{i}$ is a $d$-dimensional ($d = 10$ in this experiment) column feature vector drawn randomly from $\Nm(0,1)$ representing node $i$.
Then each entry of the common coefficient $\beta$ has a probability $p_1 = 0.4$ with nonzero value and they are drawn randomly from $\Nm(0,1)$. Besides, for each user $u$, each entry of his personalized deviation coefficient $\delta_u$ has a probability $p_2 = 0.4$ to be nonzero and
is drawn randomly from $\Nm(0,1)$. Moreover, each user has a probability $p_1 = 0.4$ having a nonzero $\gamma^u$, and those nonzero $\gamma^u$ is drawn randomly from $\N(0,2^2)$.
At last, we draw $N^u$ samples for each user randomly with binary response $y^u_{ij}$ following the model $P(y^u_{ij} = 1) = \Psi((\phi_i^T\eta+\phi_i^T\xi^u) - (\phi_j^T\eta+\phi_j^T\xi^u) + \gamma^u)$, where $\Psi(t) = 1/(1+e^{-t})$.
 The sample number $N^u$ uniformly spans in $[N_1,N_2] = [50,200]$. Finally, we obtain a multi-edge graph labeled by 100 users.

 \textbf{Comparative Results} To see whether our proposed method could provide more precise preference function for users by introducing individual-specific parameters, we randomly split the whole data sample into training set and testing set. In particular, we first split the items into training item ($75\%$ of the total items) and testing item (the remaining $25\%$). Then pairwise comparisons which contain one/two of the testing item will be pushed into the testing set, while others will be treated as training set. In other words, via this partition, for each comparisons in the testing set, at least one item is a new comer which has never appear in the training set.
 To ensure the statistical
stability, we repeat this procedure 20 times. We compare our fine-grained model with
7 competitors, i.e., RankSVM \cite{joachims2009svm}, RankBoost \cite{freund2003efficient},
RankNet \cite{burges2005learning}, gdbt\cite{GBDT}, dart\cite{dart}, Unified Robust Learning to Rank (URLR) \cite{fu2015robust},
and HodgeRank \cite{Hodge}.
Tab.\ref{tab:simu2} shows the experimental results of the proposed mixed-effects model compared with other coarse-grained models,
which indicates that all of our models exhibit smaller test error (i.e. mismatch ratio) due to their parsimonious multi-levels.
Besides, it is worth mentioning that GLM-based models (i.e. Bradley-Terry and Thurstone-Mosteller) could exhibit better
performance than linear model which suggests that these two are more suitable for binary data.

{\renewcommand\baselinestretch{1.0}\selectfont

\begin{table} [t] \renewcommand{\captionfont}{\footnotesize \bfseries}
\caption{\label{simulated}  Coarse-grained vs. fine-grained model (i.e., Ours) on test error (i.e. mismatch ratio) in simulated data.}
\centering

\newsavebox{\tablebox}

\begin{lrbox}{\tablebox}
     \begin{tabular}{lllll}
  \hline     &min  &mean &max &std\\
  \hline RankSVM{\cite{joachims2009svm}}     &0.1846 	&0.3165 	&0.4831 	&0.0750\\
 \hline RankBoost{\cite{freund2003efficient}}  &0.2216 	&0.3415 	&0.4818 	&0.0690 \\
 \hline RankNet{\cite{burges2005learning}}      &0.1980 	&0.3213 	&0.4728 	&0.0740\\
 \hline gdbt\cite{GBDT}  &0.2043 	&0.3241 	&0.4809 	&0.0767\\
\hline dart\cite{dart}    &0.2241 	&0.3235 	&0.4761 	&0.0732\\
 \hline URLR{\cite{fu2015robust}}   &0.2061	&0.3198	&0.4378	&0.0726\\
 \hline  HodgeRank\cite{Hodge}      &0.1946    &0.3097   &0.4788    &0.0760 \\
\hline  Linear      &0.1449    &\textcolor{red}{0.1892}    &0.2368    &0.0264  \\
\hline Bradley-Terry    &0.1282    &\textcolor{red}{0.1799}    &0.2355    &0.0303 \\
\hline Thurstone-Mosteller     &0.1300    &\textcolor{red}{0.1822}    &0.2368    &0.0297 \\
 \hline
 \end {tabular}
  \end{lrbox}
\scalebox{1}{\usebox{\tablebox}}
       \label{tab:simu2}
\end{table}
\par}


\textbf{Speedup of SynPar-LBI} We then demonstrate the linear speedup of the synchronized parallel LBI. In evaluating a parallel system, the typical performance measure
is \emph{speedup}, which is
defined as the ratio of the
elapsed time when executing a program on a single thread
(the single thread execution time) to the execution time
when $M$ threads are available. Let $T(M)$ be the time required to complete the task on $M$ threads. The speedup $S(M)$ is the ratio:
S(M)=T(1)/T(M).

In our setting, $M=1, 2, 3,...,16$. Fig.\ref{fig:paralbi} (Left) shows the mean running time for 20 times repeat of SynPar-LBI with thread number changing from 1 to 16 in a 16-core server with Intel(R) Xeon(R) E5-2670 2.60GHz CPU and 384GB of RAM. The server runs Linux 4.2.0 64bit. Furthermore, Fig.\ref{fig:paralbi} (Right) shows the error bar of speedup with confidence interval [0.25 0.75]. It is easy to find that the parallel LBI could speed up the running time almost in a linear manner.

\begin{figure}[t]
    \begin{minipage}{0.49\columnwidth}
    \centering
    \scriptsize
    \begin{tabular}{l|l||l|l}
        \hline\hline
        M & T(M)(s) & M  & T(M)(s) \\ \hline\hline
        1 & 232.09  & 9  & 29.38   \\ \hline
        2 & 120.67  & 10 & 26.80   \\ \hline
        3 & 83.40   & 11 & 25.27   \\ \hline
        4 & 62.70   & 12 & 24.05   \\ \hline
        5 & 51.15   & 13 & 22.55   \\ \hline
        6 & 42.04   & 14 & 20.84   \\ \hline
        7 & 37.27   & 15 & 20.23   \\ \hline
        8 & 34.60   & 16 & 20.24   \\ \hline
    \end{tabular}
    \end{minipage}
    \begin{minipage}{0.49\columnwidth}
    \centering
    \includegraphics[width=0.9\linewidth]{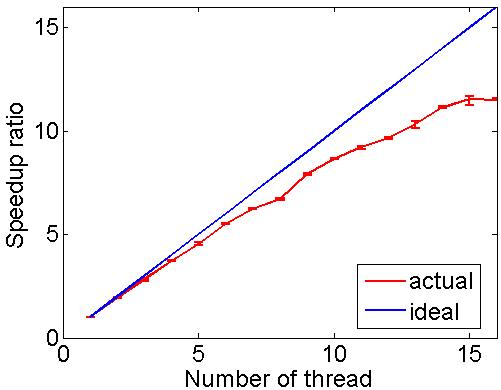}
    \end{minipage}
  \renewcommand{\captionfont}{\footnotesize \bfseries}
  \caption{Left: Mean running time (20 times repeat) of SynPar-LBI with thread number changing from 1 to 16 in simulated data. Right: The linear speedup of parallel LBI in simulated data.} \label{fig:paralbi}
\end{figure}


%
%

\subsection{Movie Preference Prediction}

\textbf{Dataset} The MovieLens 1M DataSet \footnote{https://grouplens.org/datasets/movielens/} is comprised of 3952 movies
rated by 6040 users. Each movie is rated on a scale from
1 to 5, with 5 indicating the best movie and 1 indicating
the worst movie. There are a total of one million ratings
in this dataset. Moreover, demographic information is provided voluntarily by the users, including gender, age range, occupation.
Each movie titles are identical to titles provided by the IMDb \footnote{http://www.imdb.com/} and  each can be represented as a
18-dimensional genre feature vector, including Action, Adventure, Animation, Children's, Comedy, Crime, Documentary, Drama, Fantasy, Film-Noir, Horror, Musical, Mystery, Romance, Sci-Fi, Thriller, War, Western.

\textbf{Settings} We then select a subset of this dataset containing 100 movies rated by 420 users, ensuring that each user has at least 20 ratings while each movie has been rated by at least 10 users. Since the proposed algorithm is designed for pairwise comparisons, we convert
the rating information into a set of pairwise comparisons.
More specifically, we create a pairwise comparison $(i, j)$ if item $i$ is
rated higher by user $u$ than item $j$. Note that no
pairwise comparison data is generated if two items are given
the same rating.

{\renewcommand\baselinestretch{1.0}\selectfont

\begin{table} [t] \renewcommand{\captionfont}{\footnotesize \bfseries}
\caption{\label{simulated}  Coarse-grained vs. fine-grained model (i.e., Ours) on test error (i.e. mismatch ratio) in movie dataset.}
\centering

\begin{lrbox}{\tablebox}
     \begin{tabular}{lllll}

  \hline     &min  &mean &max &std\\
  \hline RankSVM{\cite{joachims2009svm}}     &0.3580 	&0.4538 	&0.5642 	&0.0515\\
 \hline RankBoost{\cite{freund2003efficient}} &0.4249	&0.4663	&0.5140	&0.0267\\
 \hline RankNet{\cite{burges2005learning}}     &0.4190	&0.4585	&0.5168	&0.0221\\
 \hline gdbt\cite{GBDT} &0.3217 	&0.4215 	&0.5070 	&0.0453\\
\hline dart\cite{dart}   &0.2988 	&0.4335 	&0.5189 	&0.0487\\
 \hline URLR{\cite{fu2015robust}}   &0.3998 	&0.4409 	&0.4876 	&0.0230\\
 \hline  HodgeRank\cite{Hodge}     &0.3918    &0.4361    &0.4666    &0.0196 \\
\hline  Linear      &0.3103    &\textcolor{red}{0.3316}    &0.3479    &0.0105  \\
\hline Bradley-Terry     &0.3016    &\textcolor{red}{0.3278}    &0.3440    &0.0117 \\
\hline Thurstone-Mosteller     &0.3072    &\textcolor{red}{0.3291}    &0.3457    &0.0113 \\
 \hline
 \end {tabular}
  \end{lrbox}
\scalebox{1}{\usebox{\tablebox}}
       \label{tab:movie2}

\end{table}
\par}

%
%
%


\begin{figure}[t]
    \begin{minipage}{0.49\columnwidth}
    \centering
    \scriptsize
    \begin{tabular}{l|l||l|l}
        \hline\hline
        M & T(M)(s) & M  & T(M)(s) \\ \hline\hline
        1 & 629.66  & 9  & 72.61   \\ \hline
        2 & 326.75  & 10 & 64.93   \\ \hline
        3 & 220.81  & 11 & 60.74   \\ \hline
        4 & 164.83  & 12 & 54.85   \\ \hline
        5 & 134.05  & 13 & 53.22   \\ \hline
        6 & 107.34  & 14 & 48.50   \\ \hline
        7 & 94.67   & 15 & 47.91   \\ \hline
        8 & 83.79   & 16 & 43.75   \\ \hline
    \end{tabular}
    \end{minipage}
    \begin{minipage}{0.49\columnwidth}
    \centering
    \includegraphics[width=0.9\linewidth]{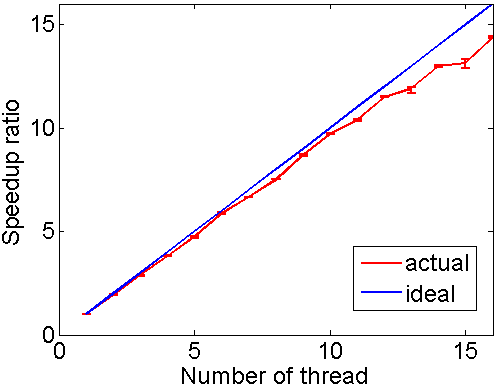}
    \end{minipage}
  \renewcommand{\captionfont}{\footnotesize \bfseries}
  \caption{Left: Mean running time (20 times repeat) of SynPar-LBI with thread number changing from 1 to 16 in movie dataset. Right: The linear speedup of parallel LBI in movie dataset.} \label{fig:movielbi}
\end{figure}

{\renewcommand\baselinestretch{1.1}\selectfont

\begin{table} [t]
\renewcommand{\captionfont}{\footnotesize \bfseries}
\caption{\label{tab:occupations} Occupations and age ranges in movie dataset.}

\makeatletter\def\@captype{table}\makeatother
\begin{minipage}[t]{0.28\textwidth}
\footnotesize
\centering

\subtable[Occupation categories]{

\begin{lrbox}{\tablebox}
       \begin{tabular}{c|l|c|l}
        \hline\hline
        Index & Occupation              & Index & Occupation             \\ \hline\hline
        1     & academic or educator    & 12    & programmer             \\
        2     & artist                  & 13    & retired                \\
        3     & clerical or admin       & 14    & sales or marketing     \\
        4     & college or grad student & 15    & scientist              \\
        5     & customer service        & 16    & self-employed          \\
        6     & doctor or health care   & 17    & technician or engineer \\
        7     & executive or managerial & 18    & tradesman or craftsman \\
        8     & farmer                  & 19    & unemployed             \\
        9     & homemaker               & 20    & writer                 \\
        10    & K-12 student            & 21    & other or not specified \\
        11    & lawyer                  & ~     & ~                      \\ \hline
        \end{tabular}
\end{lrbox}
\scalebox{0.6}{\usebox{\tablebox}}

}
\end{minipage}
\makeatletter\def\@captype{table}\makeatother
\begin{minipage}[t]{0.2\textwidth}
\scriptsize
\centering
\subtable[Age ranges]{
\begin{lrbox}{\tablebox}
       \begin{tabular}{c|c}
    \hline\hline
    Index  & Age Range  \\ \hline\hline
    1  & Under 18        \\
    2 & 18-24           \\
    3 & 25-34            \\
    4 & 35-44            \\
    5 & 45-49  \\
    6 & 50-55 \\
    7 & 56+ \\  \hline
    \end{tabular}
\end{lrbox}
\scalebox{0.6}{\usebox{\tablebox}}
\label{tab:data10-rank}

}
\end{minipage}
\end{table}
\par}

\textbf{Individual Preference} Follow the experiment design in simulated study, we also split the dataset into training set and testing set.
All the experiments were repeated 20 times to reduce variance.
Similar to the simulated dataset, the proposed fine-grained method could produce better performance than
coarse-grained models with smaller mean test error, shown in
Tab.\ref{tab:movie2}. Moreover, Fig.\ref{fig:movielbi} shows the running time of SynPar-LBI on this movie dataset and we can easily find the nearly linear speedup.

\textbf{Occupation and Age Preference}
Movie preference behavior, may be influenced by the
occupation and age factors. Tab.\ref{tab:occupations}~(a) shows the occupation categories in this dataset while Tab.\ref{tab:occupations}~(b) illustrates the age range. To exhibit the occupation influence of movie preference behavior,
users from the same occupation are treated as a group.
To further investigate the characteristics of groups with personalized preference, we plot the LBI regularization paths of the preference deviations, as has been shown in Fig.\ref{fig:moviepath}~(b) in introduction. The purple curve indicates the path of the common preference parameter, being the first popping up.
The red curves represent the top 3 groups (i.e., farmer, artist, and tradesman) who jumped out early. Groups who jumped out earlier are those with a large deviation from the common ranking.
Besides, the blue curves indicate the bottom 3 groups (i.e., homemaker, writer, and self-employed) jumped out later, and those often show similar preference with the common. In particular, the common preference is illustrated in Fig.\ref{fig:commmovie}(a) where the bars are the proportions of movie genres among top 50 movies ranked by common consensus preference.
One can see that the top four genres in the common (social) preference are Drama, Comedy, Romance, and Animation, respectively.

\begin{figure}[h]
\renewcommand{\captionfont}{\footnotesize \bfseries}
 \begin{center}
 \subfigure[]{
\includegraphics[width=0.22\textwidth]{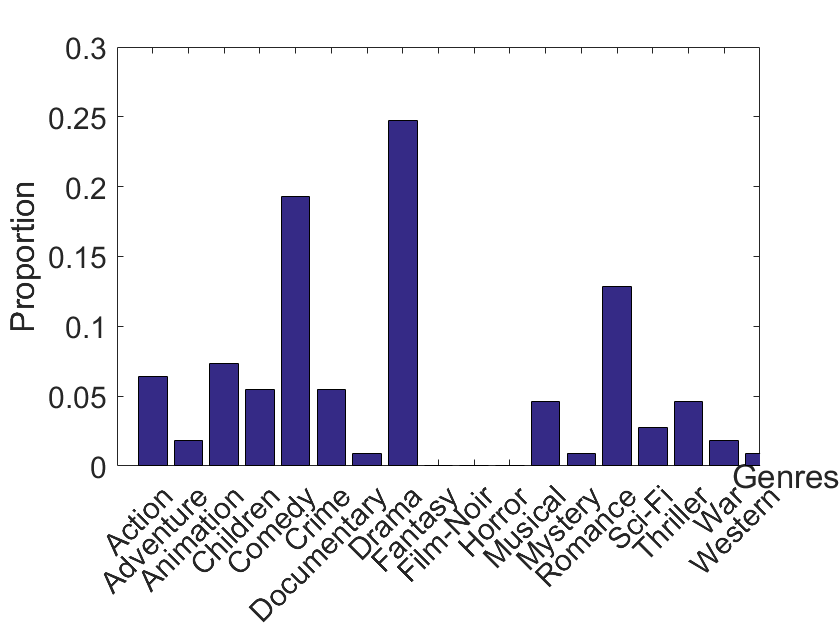}}
 \subfigure[]{
\includegraphics[width=0.2\textwidth]{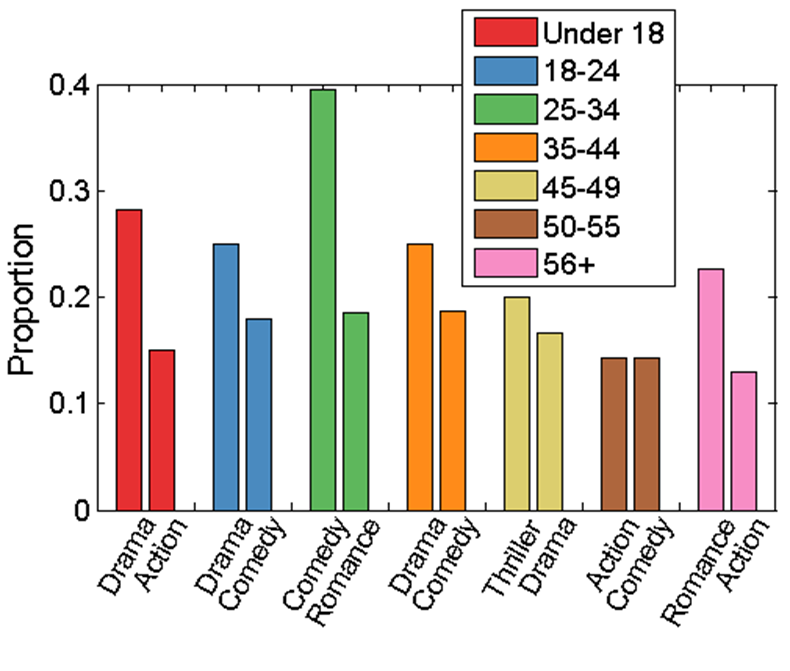}}
  \caption{(a) The common preference in MovieLens dataset; (b) Preference of 7 groups with different age range in MovieLens dataset.} \label{fig:commmovie}
\end{center}
\end{figure}

%

Despite those trends with occupation, movie preference also undergoes changes with age, and
Fig.\ref{fig:commmovie}(b) illustrates the evolution of preference over age groups. One can see that users under the year of 18 prefer
Drama and Action movies best, while
ones between 18-24 are willing to watch Drama and Comedy
instead. When users slowly waltz into their 25-34, they begin
to enjoy the love story. However, when they get to their 40s,
it happened that they grew to like the thriller movie best.
Not surprisingly, as they continue into old age such as beyond 56, their retrospect on whole life cherishes love in a deep way and Romance movie returns to
be their favourite again.

\subsection{Image Quality Assessment (IQA)}

\textbf{Settings} Two publicly available datasets, LIVE
\cite{LIVE} and IVC \cite{IVC}, are used in this work. It includes $52,043$  paired comparisons collected from 342 observers of different
cultural background. The number of responses each reference image receives is different. To validate whether the annotators' preference function we estimated is good enough, we randomly take
reference image 1 as an illustrative example while other reference
images exhibit similar results.

\begin{table}[t]
\renewcommand{\captionfont}{\footnotesize \bfseries}
\caption{\label{tab:iqa}  Coarse-grained vs. fine-grained model (i.e., Ours)  on test error (i.e. mismatch ratio) in IQA dataset (reference image 1).}

\centering
\begin{lrbox}{\tablebox}

\begin{tabular}{lllll}
 \hline     &min  &mean &max &std\\
  \hline RankSVM{\cite{joachims2009svm}}     &0.1334 	&0.1627 	&0.1808 	&0.0145\\
 \hline RankBoost{\cite{freund2003efficient}}  &0.1552  &0.1727  &0.1898	&0.0107\\
 \hline RankNet{\cite{burges2005learning}}    &0.1945 	&0.2260 	&0.2525 	&0.0136\\
 \hline gdbt\cite{GBDT} &0.1337 	&0.1491 	&0.1654 	&0.0102\\
\hline dart\cite{dart}   &0.1337 	&0.1532 	&0.1667 	&0.0113\\
 \hline URLR{\cite{fu2015robust}}   &0.1894 	&0.2140 	&0.2553 	&0.0177\\
 \hline  HodgeRank\cite{Hodge}     &0.1678    &0.1874    &0.2019    &0.0081 \\
\hline  Linear      &0.1002    &\textcolor{red}{0.1094}    &0.1239    &0.0065  \\
\hline Bradley-Terry     &0.0638    &\textcolor{red}{0.0708}    &0.0817    &0.0054 \\
\hline Thurstone-Mosteller     &0.0625     &\textcolor{red}{0.0731}    &0.0825    &0.0057  \\
 \hline
 \end {tabular}
 \end{lrbox}
\scalebox{1}{\usebox{\tablebox}}
\end{table}


\begin{figure}[t]
    \begin{minipage}{0.49\columnwidth}
    \centering
    \scriptsize
    \begin{tabular}{l|l||l|l}
        \hline\hline
        M & T(M)(s) & M  & T(M)(s) \\ \hline\hline
        1 & 57.38   & 9  & 7.54    \\ \hline
        2 & 29.98   & 10 & 7.00    \\ \hline
        3 & 20.99   & 11 & 6.40    \\ \hline
        4 & 16.11   & 12 & 6.11    \\ \hline
        5 & 13.06   & 13 & 5.97    \\ \hline
        6 & 11.16   & 14 & 5.76    \\ \hline
        7 & 9.69    & 15 & 5.53    \\ \hline
        8 & 8.82    & 16 & 5.29    \\ \hline
    \end{tabular}
    \end{minipage}
    \begin{minipage}{0.49\columnwidth}
    \centering
    \includegraphics[width=0.9\linewidth]{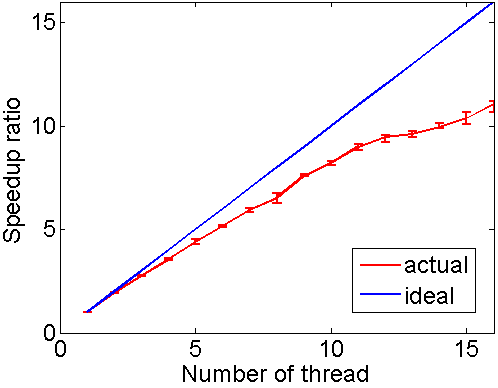}
    \end{minipage}
  \renewcommand{\captionfont}{\footnotesize \bfseries}
  \caption{Left: Mean running time (20 times repeat) of SynPar-LBI with thread number changing from 1 to 16 in IQA dataset. Right: The linear speedup of parallel LBI in IQA dataset.} \label{fig:iqalbi}
\end{figure}

\begin{figure}
\renewcommand{\captionfont}{\footnotesize \bfseries}
 \begin{center}
  \subfigure[Linear]{
\includegraphics[width=0.15\textwidth]{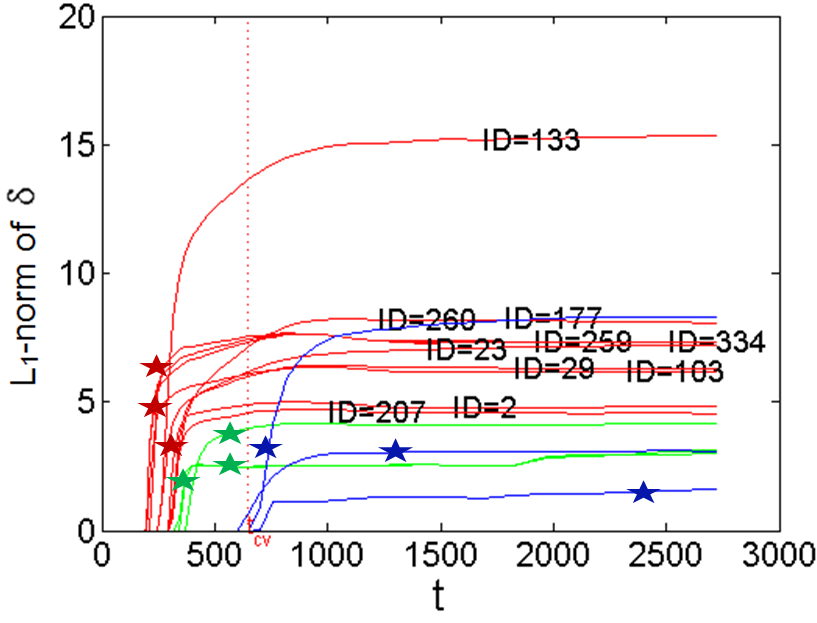}}
   \subfigure[BT]{
\includegraphics[width=0.15\textwidth]{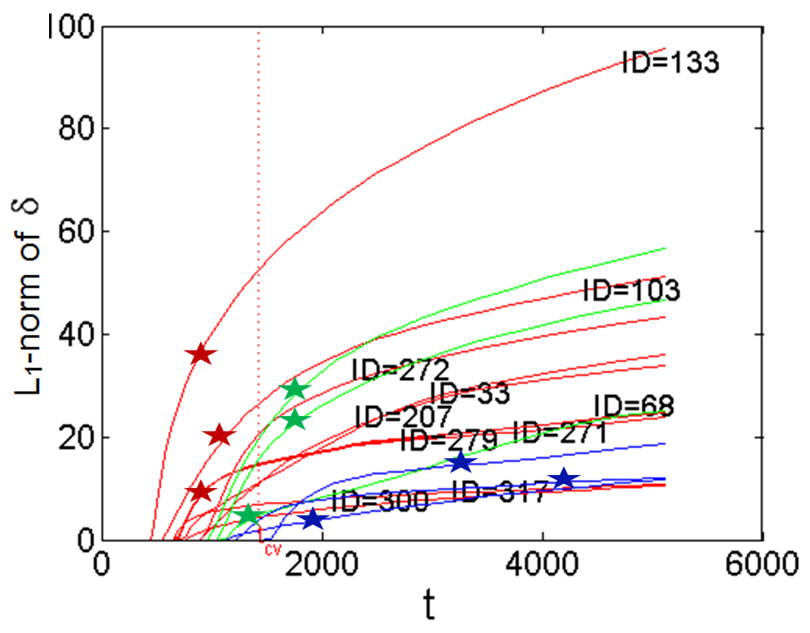}}
\subfigure[TM]{
\includegraphics[width=0.15\textwidth]{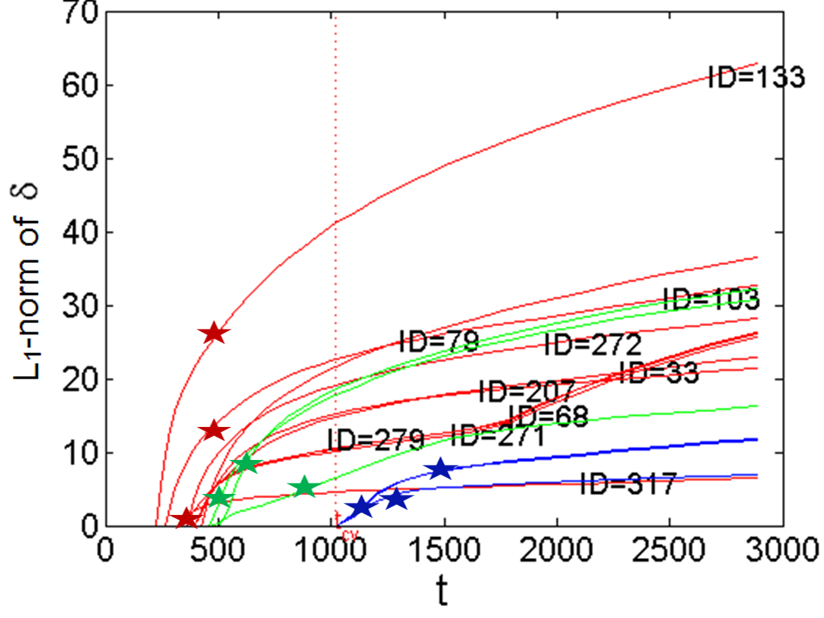}}
  \caption{LBI regularization path of $\delta$ exhibiting personalized ranking in IQA dataset (reference image 1). (Red: top 10 personalized ranking annotators; Green: middle 3; Blue: bottom 3)}\label{fig:iqaref1preferencepath}
\end{center}
\end{figure}

\begin{figure}
\renewcommand{\captionfont}{\footnotesize \bfseries}
 \begin{center}
  \subfigure[Linear]{
\includegraphics[width=0.14\textwidth]{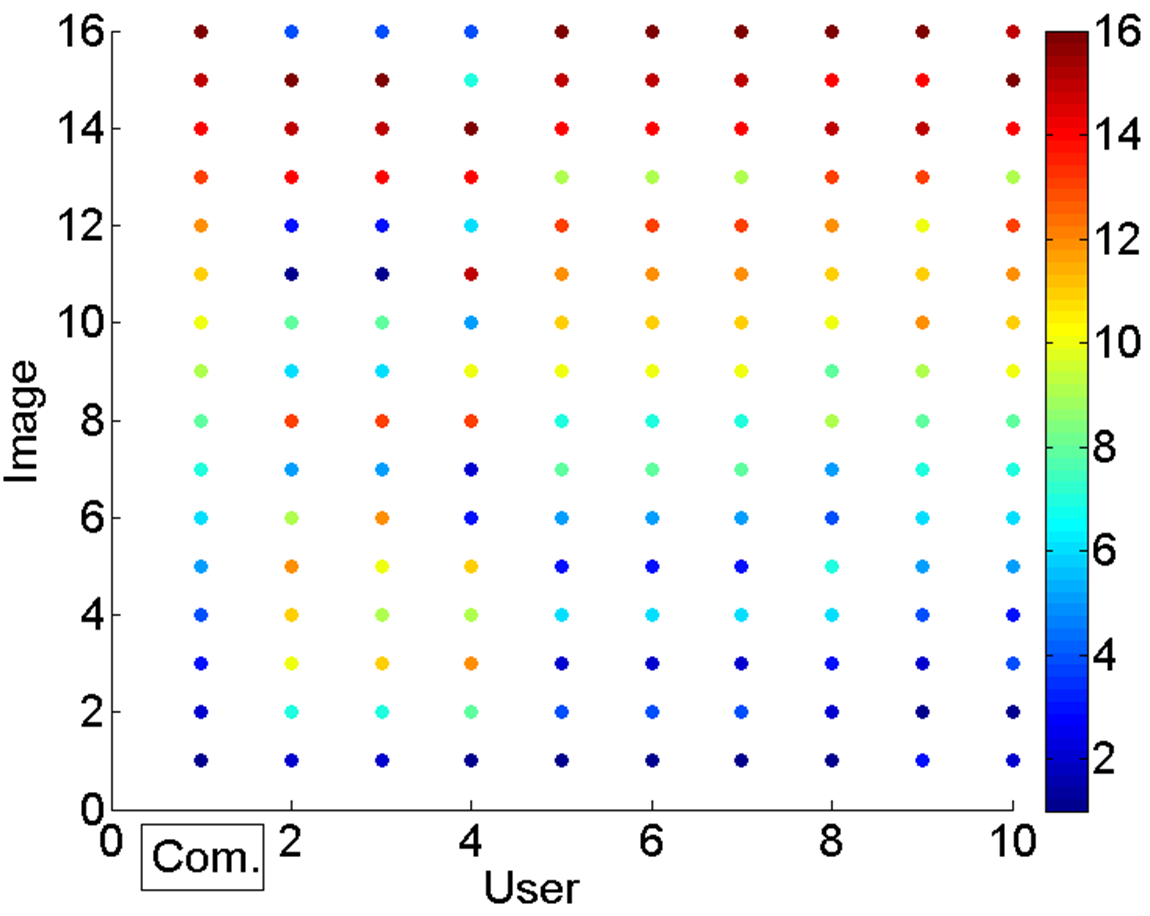}}
   \subfigure[BT]{
\includegraphics[width=0.14\textwidth]{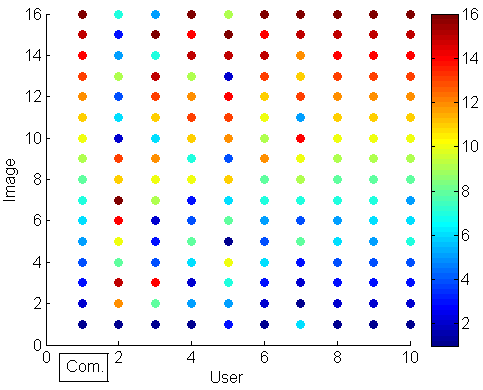}}
\subfigure[TM]{
\includegraphics[width=0.14\textwidth]{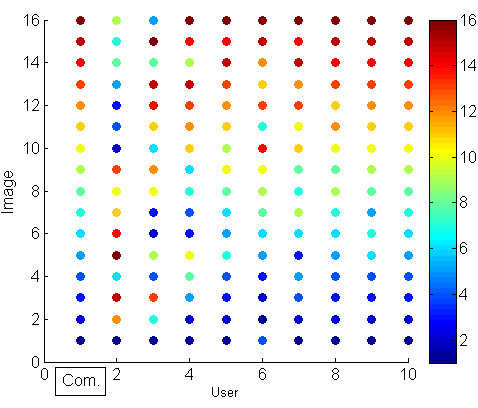}}
  \caption{Ranking order comparison of common vs. personalized rankings of 9 representative annotators in IQA dataset (reference image 1).}  \label{iqa_position_color}
\end{center}
\end{figure}

\begin{figure}
\renewcommand{\captionfont}{\footnotesize \bfseries}
 \begin{center}
  \subfigure[Linear]{
\includegraphics[width=0.15\textwidth]{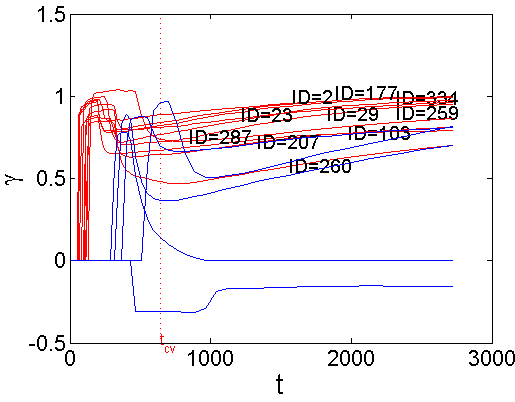}}
   \subfigure[BT]{
\includegraphics[width=0.15\textwidth]{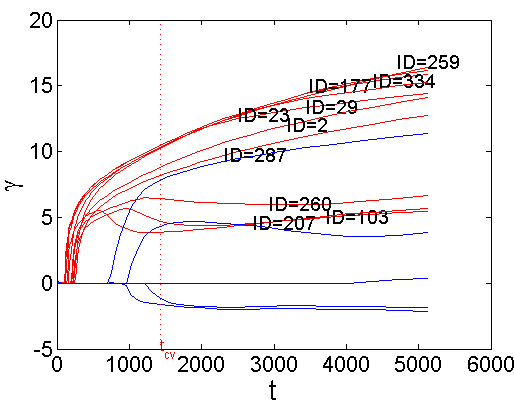}}
\subfigure[TM]{
\includegraphics[width=0.15\textwidth]{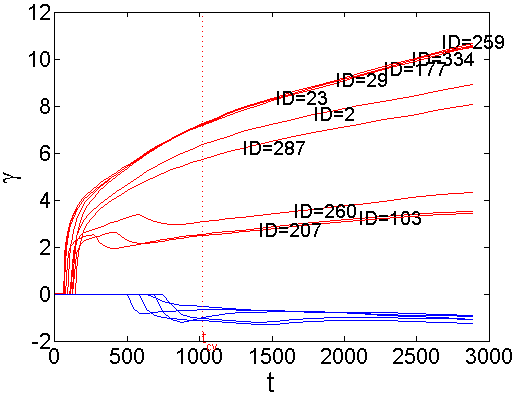}}
  \caption{LBI regularization path of $\gamma$ exhibiting position bias in IQA dataset (reference image 1). (Red: top 10 position-biased annotators; Blue: bottom 5 position-biased annotators).} \label{fig:iqa_position}
\end{center}
\end{figure}


\textbf{Results}  Tab.\ref{tab:iqa} shows the mean test error (70\% data for training, 30\% for testing)
results of 20 times achieved by this scheme. It is shown that consistent with the simulated data, in this dataset, the mixed-effects model with three losses could also provide
better approximate results of the annotators' preference than the HodgeRank estimator. Moreover, Fig.\ref{fig:iqalbi} shows the running time of SynPar-LBI on this IQA dataset and we can easily find the nearly linear speedup.

 To further investigate the characteristics of annotators with personalized ranking,  Fig.\ref{fig:iqaref1preferencepath} illustrates annotator's LBI regularization paths of preference deviations with optimal $t$ (i.e., $t_{cv}$) returned by cross-validation in three losses. The red curves in Fig.\ref{fig:iqaref1preferencepath} represent the top 10 annotators who jumped out early. Moreover, Fig.\ref{iqa_position_color} shows the order comparisons of common ranking (i.e., com.) and personalized ranking of 9 representative annotators at $t_{cv}$. The X-axis represents user index: user = 2, 3, 4 jumped out early corresponding to paths labeled with red stars in Fig.\ref{fig:iqaref1preferencepath}; user = 5, 6, 7 jumped out in the middle time corresponding to green stars; user = 8, 9, 10 jumped out late corresponding to blue stars. The order of faces in Y-axis is arranged from lower to higher (i.e., from color blue to red) according to the common ranking score calculated by our method. The color represents the ranking position returned by the corresponding user. It is easy to see users jumped out late exhibit almost consistent ranking order with the common ranking, while the earlier ones are almost the adversarial against the common.

\textbf{Remark} It is easy to see
that among the top 10 annotators returned by linear model, 9 of them (except annotator
with ID = 133) click one side almost all the time (i.e., position-biased
annotators), while results returned by other two are not. The reason of such a phenomenon lies in the difference of linear model and probability model. Such kind of user always has $y^u_{ij}\equiv 1$ or $-1$. In simple linear model, to fit such user, only a position bias term $\gamma^u$ is not enough. Since the common score $\eta$ always exists and is nonzero, only $\gamma^u = 1$ or $-1$ and $\xi^u = -\eta$ can fit the data well, so under the linear model, these users' $\xi$ is nonzero. While in the other two GLM, the probability explain makes a single $\gamma^u$ enough to fit the data. Since a $\gamma^u$ with much larger magnitude the $\eta_i-\eta_j$ can already dominant the probability $\Psi((\eta_i+\xi_i^u) - (\eta_j+\xi_j^u) + \gamma^u)$ even $\xi^u = 0$. Also in the other data, there also exist such kind of users, but their samples are not as many as those in this data, so such a phenomenon is not observed in the other data. This gives an example that GLM can be qualitatively better than the linear model for binary comparison data.

{\renewcommand\baselinestretch{1.1}\selectfont
\setlength{\belowcaptionskip}{0pt}
\begin{table}[t]
\renewcommand{\captionfont}{\footnotesize \bfseries}
\caption{\label{tab:ref1} Top 10 position-biased annotators in IQA dataset (reference image 1).}
\tiny
\centering
\begin{lrbox}{\tablebox}
\begin{tabular}{||c|c|c|c||c|c|c|c||}
  \hline  \textbf{Order} &\textbf{ID}   &\textbf{Left}  &\textbf{Right} & \textbf{Order} & \textbf{ID}   &\textbf{Left}  &\textbf{Right} \\
 \hline
 \hline  1 & \textcolor{blue}{\textbf{259}}    &96     &0  & 6 & \textcolor{blue}{\textbf{2}}   &55     &0 \\
\hline  2 & \textcolor{blue}{\textbf{334}}     &90    & 0  & 7 & \textbf{260}    & 49     &2\\
 \hline  3 & \textcolor{blue}{\textbf{177}}   &77     &0   & 8 & \textcolor{blue}{\textbf{23}}    &42     &0 \\
 \hline 4 & \textbf{103}  &74     &4  &9  & \textbf{207}    &46     &2\\
 \hline 5 & \textcolor{blue}{\textbf{29}}  &58     &0  & 10 & \textcolor{blue}{\textbf{287}}   &34     &0\\

 \hline
\end {tabular}
\medskip
\end{lrbox}
\scalebox{1}{\usebox{\tablebox}}
\end{table}
\par}

Moreover, Fig.\ref{fig:iqa_position} illustrates the LBI regularization paths of annotator's position bias with red lines represent the top 10 annotators. It is easy to see that the corresponding results returned from these three loss functions are exactly the same. Tab.\ref{tab:ref1} further shows the click counts of each side (i.e., Left and Right) for these top 10 position-biased annotators.
It is easy to see that these annotators can be divided
into two types: (1) click one side all the time (with
ID in blue); (2) click one side with high probability
(others). Although it
might be relatively easy to identify the annotators of type (1) above
by inspecting their inputs, it is impossible for eye inspection
to pick up those annotators of type (2) with mixed rational and abnormal behaviors.
Therefore it is essential to design such a statistical methodology to
quantitatively detect these kind of position-biased annotators
for crowdsourcing platforms in market. It is interesting to see that annotators highlighted with blue color in
Tab.\ref{tab:ref1} click the left side all the time. We then go back to
the crowdsourcing platform and find out that the reason behind
this is a \emph{default choice} on the left button, which induces
some lazy annotators to cheat for the task.

\begin{table}[h]
\renewcommand{\captionfont}{\footnotesize \bfseries}
\caption{\label{tab:university}  Coarse-grained vs. fine-grained model (i.e.,
Ours) on test error (i.e. mismatch ratio) in WorldCollege ranking dataset.}
\centering
\begin{lrbox}{\tablebox}
\begin{tabular}{lllll}
 \hline     &min  &mean &max &std\\
  \hline RankSVM{\cite{joachims2009svm}}     &0.3009 	&0.3165 	&0.3289 	&0.0068\\
 \hline RankBoost{\cite{freund2003efficient}} 	&0.3199	&0.3321	&0.3448	&0.0073\\
 \hline RankNet{\cite{burges2005learning}}    &0.3393 	&0.3533 	&0.3710 	&0.3533\\
 \hline gdbt\cite{GBDT} 	&0.3436 	&0.3579 	&0.3730 	&0.0086 \\
\hline dart\cite{dart}   &0.3470 	&0.3716 &0.3986 	&0.0134\\
 \hline URLR{\cite{fu2015robust}}  &0.2893 	&0.3136 	&0.3269 	&0.0075\\
 \hline  HodgeRank\cite{Hodge}     &0.2979    &0.3108    &0.3230    &0.0097  \\
\hline  Linear      &0.2530    &\textcolor{red}{0.2670}    &0.2757    &0.0071  \\
\hline Bradley-Terry     &0.2456    &\textcolor{red}{0.2555}    &0.2678    &0.0099 \\
\hline Thurstone-Mosteller      &0.2553    &\textcolor{red}{0.2616}    &0.2696    &0.0067  \\
 \hline
 \end {tabular}
 \end{lrbox}
\scalebox{1}{\usebox{\tablebox}}

\end{table}


\begin{figure}[t]
    \begin{minipage}{0.49\columnwidth}
    \centering
    \scriptsize
    \begin{tabular}{l|l||l|l}
        \hline\hline
        M & T(M)(s) & M  & T(M)(s) \\ \hline\hline
        1 & 221.07  & 9  & 25.65   \\ \hline
        2 & 114.99  & 10 & 22.90   \\ \hline
        3 & 78.40   & 11 & 21.42   \\ \hline
        4 & 58.09   & 12 & 19.69   \\ \hline
        5 & 47.80   & 13 & 18.92   \\ \hline
        6 & 38.21   & 14 & 17.55   \\ \hline
        7 & 33.54   & 15 & 16.83   \\ \hline
        8 & 29.60   & 16 & 15.51   \\ \hline
    \end{tabular}
    \end{minipage}
    \begin{minipage}{0.49\columnwidth}
    \centering
    \includegraphics[width=0.9\linewidth]{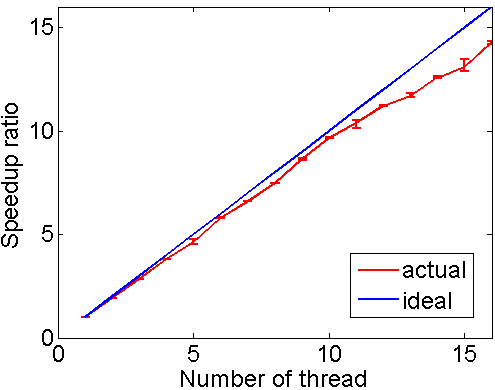}
    \end{minipage}
  \renewcommand{\captionfont}{\footnotesize \bfseries}
  \caption{Left: Mean running time (20 times repeat) of SynPar-LBI with thread number changing from 1 to 16 in WorldCollege ranking dataset. Right: The linear speedup of parallel LBI in WorldCollege ranking dataset.} \label{fig:univerlbi}
\end{figure}

\begin{figure}
\renewcommand{\captionfont}{\footnotesize \bfseries}
 \begin{center}
  \subfigure[Linear]{
\includegraphics[width=0.14\textwidth]{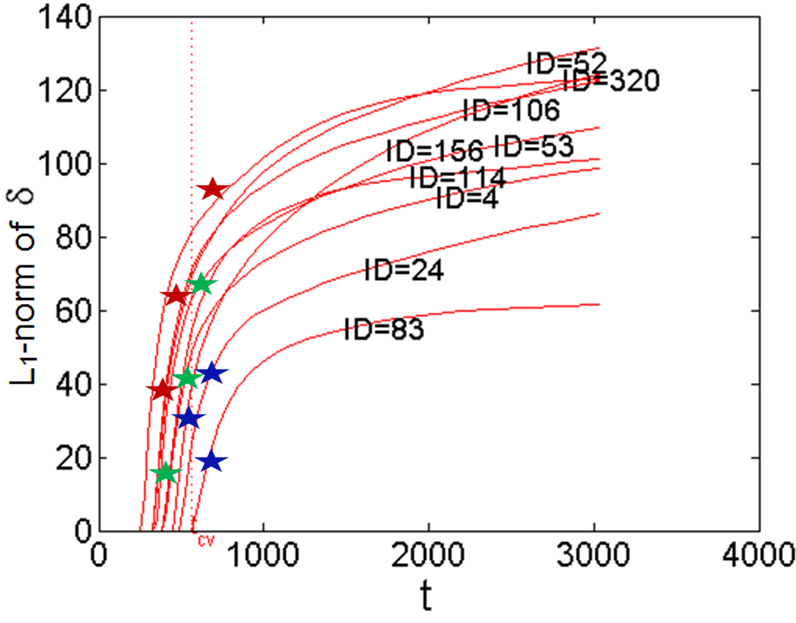}\label{fig:univer_l2_preference}}
   \subfigure[BT]{
\includegraphics[width=0.14\textwidth]{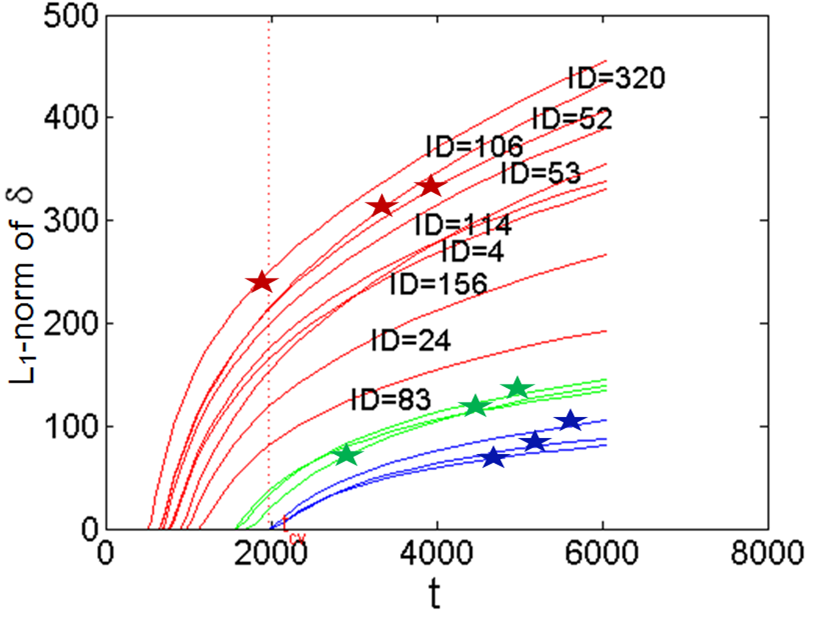}\label{fig:univer_model1_preference}}
\subfigure[TM]{
\includegraphics[width=0.14\textwidth]{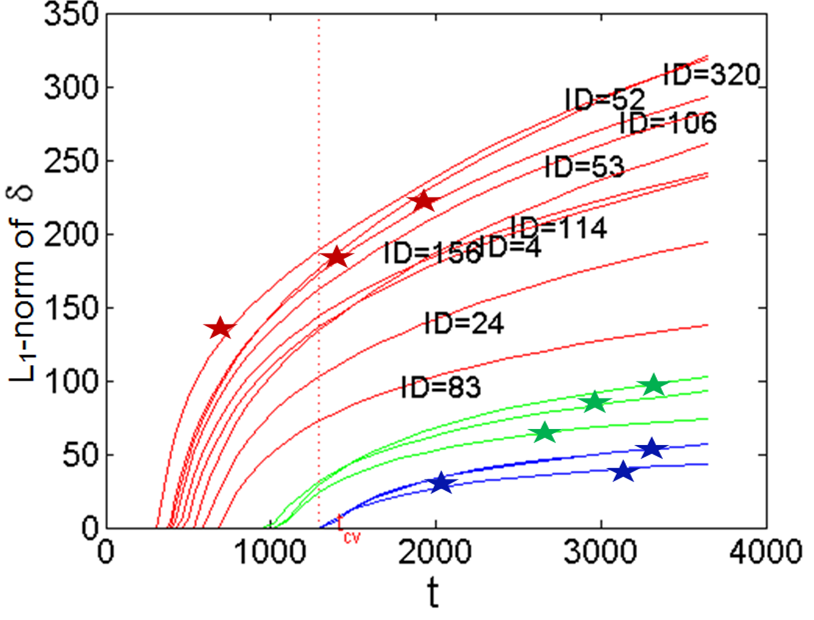}\label{fig:univer_model2_preference}}
  \caption{LBI regularization path of $\delta$ exhibiting personalized ranking in WorldCollege ranking dataset. (Red: top 9 personalized ranking annotators; Green: middle 3; Blue: bottom 3)}
\end{center}
\end{figure}

\begin{figure}
\renewcommand{\captionfont}{\footnotesize \bfseries}
 \begin{center}
  \subfigure[Linear]{
\includegraphics[width=0.15\textwidth]{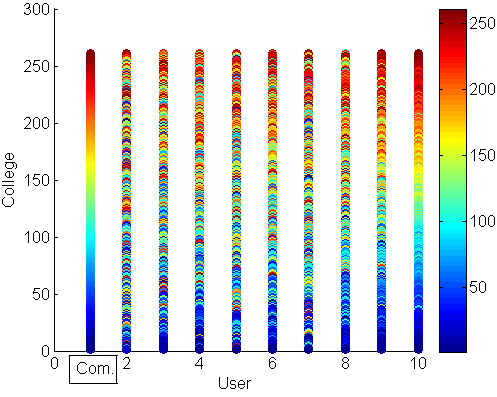}}
   \subfigure[BT]{
\includegraphics[width=0.15\textwidth]{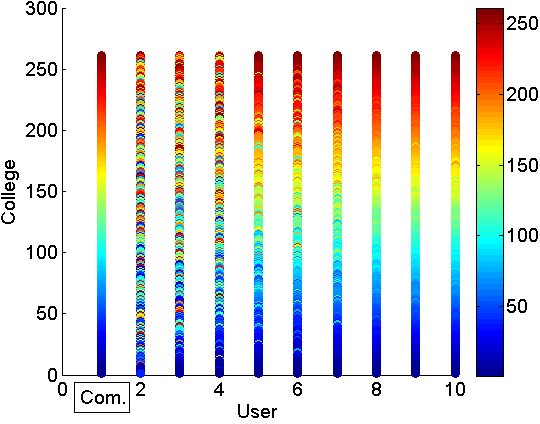}}
\subfigure[TM]{
\includegraphics[width=0.15\textwidth]{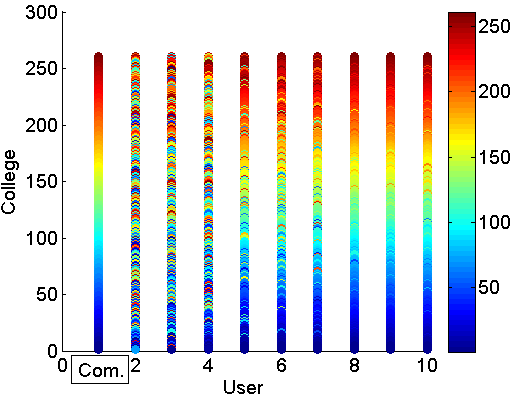}}
 \caption{Ranking order comparison of common vs. personalized rankings of 9 annotators in WorldCollege ranking dataset.} \label{university_position_color}
\end{center}
\end{figure}


\begin{figure}[htb]
\renewcommand{\captionfont}{\footnotesize \bfseries}
 \begin{center}
  \subfigure[Linear]{
\includegraphics[width=0.15\textwidth]{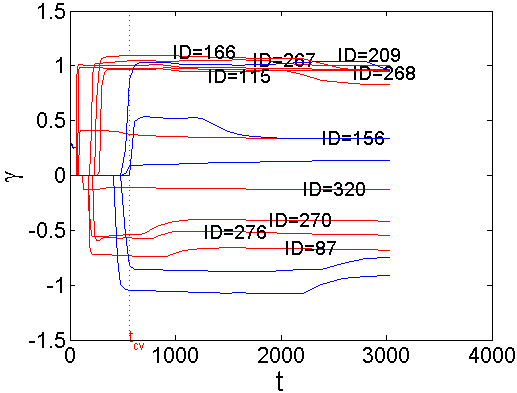}}
   \subfigure[BT]{
\includegraphics[width=0.15\textwidth]{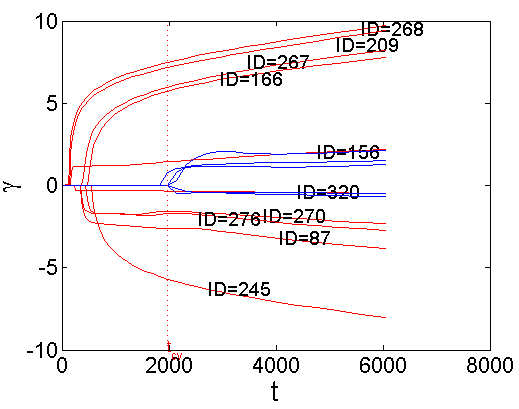}}
\subfigure[TM]{
\includegraphics[width=0.15\textwidth]{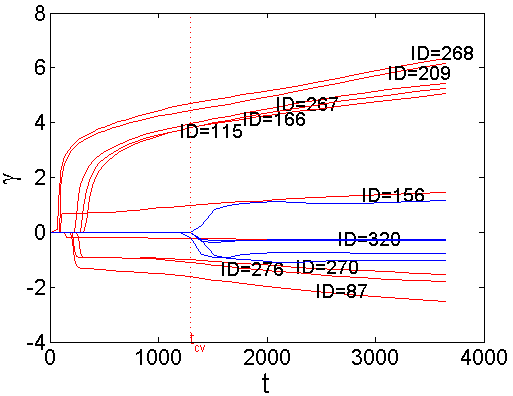}}
  \caption{LBI regularization path of $\gamma$ exhibiting position bias in WorldCollege ranking dataset. (Red: top 10 position-biased annotators; Blue: bottom 5 position-biased annotators).} \label{fig:university_position}
\end{center}
\end{figure}

\subsection{WorldCollege Ranking}

\textbf{Settings} We now apply the proposed method to the WorldCollege dataset, which is composed of 261 colleges. Using the \href{http://www.allourideas.org/}{Allourideas} crowdsourcing platform, a total of 340 distinct annotators from various countries (e.g., USA, Canada, Spain, France, Japan)
are shown randomly with pairs of these colleges, and asked to decide which of the
two universities is more attractive to attend. Finally, we obtain a
total of 8,823 pairwise comparisons.

\textbf{Results} We apply the proposed method to
the resulting dataset and find out that, similar to the simulation and other two real-world datasets, the mixed-effects model could produce better performance than Hodgerank with smaller mean test error, shown in Tab.\ref{tab:university}. Moreover, Fig.\ref{fig:univerlbi} shows the linear speedup of SynPar-LBI on this dataset.
Besides, noting in this dataset, only 9 annotators are treated as annotators with distinct personalized rankings at optimal $t$ (i.e., $t_{cv}$) selected via cross-validation in linear model case, as is shown in Fig.\ref{fig:univer_l2_preference}.
However, other two losses with smaller mean test error detect more than 9 personalized-ranking annotators via cross-validation.
To better illustrate comparison result of three losses, for the other two, we also only show the top 9 annotators, as is shown in Fig.\ref{fig:univer_model1_preference} and \ref{fig:univer_model2_preference}. It is pleasing to see that the top 9 annotators returned by three losses are exactly the same in this dataset.
The common ranking vs. personalized ranking of 9 representative users is shown in Fig.\ref{university_position_color} with a similar observation to the other two datasets.
Besides, the regularization paths of position bias and click counts of top 10 annotators in this dataset are shown in Fig.\ref{fig:university_position} and Tab.\ref{tab:universityposition}.
It is easy to see that similar to the human age dataset, these annotators are either clicking one
side all the time, or clicking one side with high probability in mixed behaviors. Clearly, when showing top 10 position-biased annotators, there is only one difference among these three cases, where linear model and Thurstone-Mosteller both pick out annotator with ID=115, while Bradley-Terry treats annotator with ID=245 as position-biased one.
A further inspection of the dataset
confirms that such a detection result is reasonable, as the ratio of left/right clicks of these two annotators are 34:0 and 0:34 respectively, as is shown in Tab.\ref{tab:universityposition}.

{\renewcommand\baselinestretch{1.1}\selectfont
\setlength{\belowcaptionskip}{0pt}
\begin{table}[t]
\renewcommand{\captionfont}{\footnotesize \bfseries}
\caption{\label{tab:universityposition} Top 10 position-biased annotators in WorldCollege ranking dataset.}
\tiny
\centering
\begin{lrbox}{\tablebox}
\begin{tabular}{||c|c|c|c||c|c|c|c||}
  \hline  \textbf{Order} &\textbf{ID}   &\textbf{Left}  &\textbf{Right} & \textbf{Order} & \textbf{ID}   &\textbf{Left}  &\textbf{Right} \\
 \hline
 \hline  1 & \textcolor{blue}{\textbf{268}} &148	&0 & 6 & \textbf{270}	&20	&70 \\
 \hline  2 & \textcolor{blue}{\textbf{209}}	&127	&0 & 7 & \textcolor{blue}{\textbf{267}}	&45	&0 \\
 \hline 3 & \textbf{156}	&189	&67 & 8 & \textbf{276}	&16	&54 \\
 \hline 4  & \textbf{320}	&253	&324  & 9 & \textcolor{blue}{\textbf{166}}	&35	&0 \\
 \hline 5 & \textbf{87}	&11	&62 & 10 & \textcolor{blue}{\textbf{115}}	&34	&0 \\
 \hline   &  	& 	&  & 10 & \textcolor{blue}{\textbf{245}}	&0	&34 \\

 \hline
\end {tabular}
\medskip
\end{lrbox}
\scalebox{1}{\usebox{\tablebox}}
\end{table}
\par}

\section{Conclusions}\label{sec:conclusion}

In this paper, we propose a parsimonious mixed-effects model based on HodgeRank to learn user's preference or utility function in crowdsourced ranking, which
takes into account both the personalized preference deviations from the common and position biases of the annotators.
To be specific, common preference scores indicate the consistent ranking on population-level which approximates the behavior of
all users, while a small set of annotators might have nonzero
personalized deviations and abnormal behavior in position bias. Equipped with the newly developed Linearized
Bregman Iteration, which is a simple iterative procedure generating a sequence of parsimonious models,  we establish a dynamic path from the common utility to individual
variations, with different levels of parsimony or sparsity on personalization. In this dynamic scheme, three kinds of models are systematically discussed, including the linear model with L2 loss, the Bradley-Terry model, and the Thurstone-Mosteller model.
Experimental studies conducted on simulated examples and real-world
datasets show that our proposed method could exhibit better performance (i.e.
smaller test error) compared with the traditional HodgeRank. In addition, generalized linear models may be more efficient to fit binary comparison data in terms of both the reduction of pairwise mismatch (Kendall $\tau$-distance) from observations and the discrimination of position bias from personalized preference deviations.
Our results suggest that the proposed methodology is an effective tool to investigate the diversity in annotator's behavior
in modern crowdsourced preference data.

\section{Acknowledgments}

The research of Qianqian Xu was supported in part by National Key Research and Development Plan (No.2016YFB0800403), in part by National Natural Science Foundation of China (No.61672514, 61390514, 61572042), Beijing Natural Science Foundation (4182079), Youth Innovation Promotion Association CAS, and CCF-Tencent Open Research Fund. The research of Xiaochun Cao was supported in part by National Natural Science Foundation of China (No.U1636214, 61650202), Beijing Natural Science Foundation (No.4172068), Key Program of the Chinese Academy of Sciences (No.QYZDB-SSW-JSC003).
The research of Qingming Huang was supported in part by National Natural Science Foundation of China: 61332016, 61620106009, U1636214 and 61650202, in part by National Basic Research Program of China (973 Program): 2015CB351800, in part by Key Research Program of Frontier Sciences, CAS: QYZDJ-SSW-SYS013. The research of Yuan Yao was supported in part by Hong Kong Research Grant Council (HKRGC) grant 16303817, National Basic Research Program of
China (No. 2015CB85600, 2012CB825501), National Natural Science Foundation of China (No. 61370004, 11421110001), as well as awards
from Tencent AI Lab, Si Family Foundation, Baidu Big Data Institute, and Microsoft Research-Asia.

%
%

\ifCLASSOPTIONcaptionsoff
  \newpage
\fi

\bibliographystyle{abbrv}
 \bibliography{sigproc}

\vspace{0cm}\begin{biography}[{\includegraphics[width=1in,height=1.25in,clip,keepaspectratio]{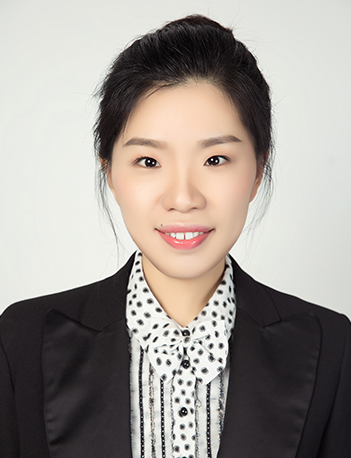}}]{Qianqian Xu received the B.S. degree in computer science from China University of Mining and Technology in 2007 and the Ph.D. degree in computer science from University of Chinese Academy of Sciences in 2013. She is currently an Associate Professor with the Institute of Computing Technology, Chinese Academy of Sciences, Beijing, China. Her research interests include statistical machine learning, with applications in multimedia and computer vision. She has authored or coauthored 10+ academic papers in prestigious international journals and conferences, among which she has published 5 full papers with the first author's identity in ACM Multimedia.  She served as member of professional committee of CAAI, and member of online program committee of VALSE, etc.}
\end{biography}

\vspace{0cm}\begin{biography}[{\includegraphics[width=1in,height=1.25in,clip,keepaspectratio]{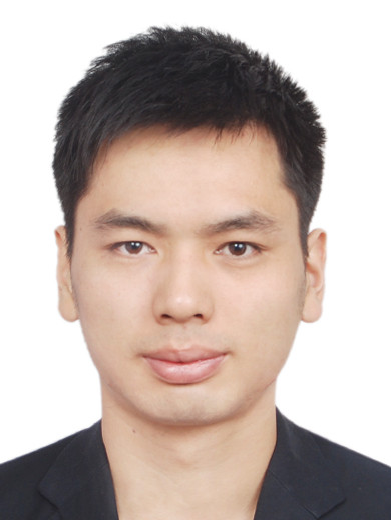}}]{Jiechao Xiong received the B.S. degree and the Ph.D. degree in statistics from the School of Mathematical Science, Peking University, Beijing, China, in 2011 and 2016, respectively. He is currently with Tencent AI Laboratory, Shenzhen, China. His research interests include statistical machine learning, algorithms, and artificial intelligence with industrial applications.}
\end{biography}

\vspace{0cm}\begin{biography}[{\includegraphics[width=1in,height=1.25in,clip,keepaspectratio]{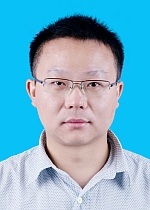}}]{Xiaochun Cao, Professor of the Institute of Information Engineering, Chinese Academy of Sciences. He received the B.E. and M.E. degrees both in computer science from Beihang University (BUAA), China, and the Ph.D. degree in computer science from the University of Central Florida, USA, with his dissertation nominated for the university level Outstanding Dissertation Award. After graduation, he spent about three years at ObjectVideo Inc. as a Research Scientist. From 2008 to 2012, he was a professor at Tianjin University. He has authored and coauthored over 100 journal and conference papers. In 2004 and 2010, he was the recipients of the Piero Zamperoni best student paper award at the International Conference on Pattern Recognition. He is a fellow of IET and a Senior Member of IEEE. He is an associate editor of IEEE Transactions on Image Processing and IEEE Transactions on Circuits and Systems for Video Technology.}
\end{biography}

\vspace{0cm}\begin{biography}[{\includegraphics[width=1in,height=1.25in,clip,keepaspectratio]{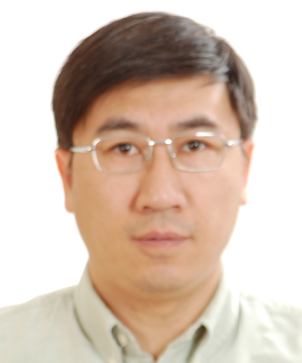}}]{Qingming Huang is a professor in the University of Chinese Academy of Sciences and an adjunct research professor in the Institute of Computing Technology, Chinese Academy of Sciences. He graduated with a Bachelor degree in Computer Science in 1988 and Ph.D. degree in Computer Engineering in 1994, both from Harbin Institute of Technology, China. His research areas include multimedia computing, image processing, computer vision and pattern recognition. He has authored or coauthored more than 300 academic papers in prestigious international journals and top-level international conferences. He is the associate editor of IEEE Trans. on CSVT and Acta Automatica Sinica, and the reviewer of various international journals including IEEE Trans. on PAMI, IEEE Trans. on Image Processing, IEEE Trans. on Multimedia, etc. He is a Fellow of IEEE and has served as general chair, program chair, track chair and TPC member for various conferences, including ACM Multimedia, CVPR, ICCV, ICME, ICMR, PCM, BigMM, PSIVT, etc.}
\end{biography}

\vspace{0cm}\begin{biography}[{\includegraphics[width=1in,height=1.25in,clip,keepaspectratio]{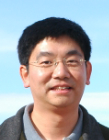}}]{Yuan Yao received the B.S.E and M.S.E in control engineering both from Harbin Institute of Technology, China, in 1996 and 1998, respectively, M.Phil in mathematics from City University of Hong Kong in 2002, and Ph.D. in mathematics from the University of California, Berkeley, in 2006. Since then he has been with Stanford University and in 2009, he joined the Department of Probability and Statistics in School of Mathematical Sciences, Peking University, Beijing, China. He is currently an Associate Professor of Mathematics, Chemical \& Biological Engineering, and by courtesy, Computer Science \& Engineering, Hong Kong University of Science and Technology, Clear Water Bay, Kowloon, Hong Kong SAR, China. His current research interests include topological and geometric methods for high dimensional data analysis and statistical machine learning, with applications in computational biology, computer vision, and information retrieval. Dr. Yao is a member of American Mathematical Society (AMS), Association for Computing Machinery (ACM), Institute of Mathematical Statistics (IMS), and Society for Industrial and Applied Mathematics (SIAM). He served as area or session chair in NIPS and ICIAM, as well as a reviewer of Foundation of Computational Mathematics, IEEE Trans. Information Theory, J.  Machine Learning Research, and Neural Computation, etc.}
\end{biography}

\end{document}